# Abundance, distribution, and origin of $^{60}$Fe in the solar protoplanetary disk

Haolan Tang,[1*] Nicolas Dauphas[1]

[1]Origins Laboratory, Department of the Geophysical Sciences and Enrico Fermi Institute, The University of Chicago, 5734 South Ellis Avenue, Chicago IL 60637.
[*]To whom correspondence should be addressed. E-mail: cafetang@uchicago.edu

**Abstract:**

Meteorites contain relict decay products of short-lived radionuclides that were present in the protoplanetary disk when asteroids and planets formed. Several studies reported a high abundance of $^{60}$Fe ($t_{1/2}$=2.62±0.04 Myr) in chondrites ($^{60}$Fe/$^{56}$Fe~6×10$^{-7}$), suggesting that planetary materials incorporated fresh products of stellar nucleosynthesis ejected by one or several massive stars that exploded in the vicinity of the newborn Sun. We measured $^{58}$Fe/$^{54}$Fe and $^{60}$Ni/$^{58}$Ni isotope ratios in whole rocks and constituents of differentiated achondrites (ureilites, aubrites, HEDs, and angrites), unequilibrated ordinary chondrites Semarkona (LL3.0) and NWA 5717 (ungrouped petrologic type 3.05), metal-rich carbonaceous chondrite Gujba (CBa), and several other meteorites (CV, EL H, LL chondrites; IIIAB, IVA, IVB iron meteorites). We derive from these measurements a much lower initial $^{60}$Fe/$^{56}$Fe ratio of (11.5±2.6)×10$^{-9}$ and conclude that $^{60}$Fe was homogeneously distributed among planetary bodies. This low ratio is consistent with derivation of $^{60}$Fe from galactic background ($^{60}$Fe/$^{56}$Fe≈2.8×10$^{-7}$ in the interstellar medium from γ-ray observations) and can be reconciled with high $^{26}$Al/$^{27}$Al~5×10$^{-5}$ in chondrites if solar material was contaminated through winds by outer layers of one or several massive stars (*e.g.*, a Wolf-Rayet star) rich in $^{26}$Al and poor in $^{60}$Fe. We present the first chronological application of the $^{60}$Fe-$^{60}$Ni decay system to establish the time of core formation on Vesta at $3.7^{+2.5}_{-1.7}$ Myr after condensation of calcium-aluminum-rich inclusions (CAIs).

## 1. Introduction

Extinct radionuclides have proven to be extremely useful for studying the astrophysical context of solar system formation, the relative chronology of early solar system events, the irradiation history of early formed solids, the timescale of nucleosynthetic processes, and the thermal history of planetary bodies (see recent reviews by Meyer and Clayton, 2000; McKeegan and Davis, 2004; Wadhwa et al., 2006; Wasserburg et al. 2006; Dauphas and Chaussidon, 2011). Many of these extinct radionuclides originated from the long-term chemical evolution of the Galaxy (*e.g.*, $^{92}$Nb, $^{129}$I, $^{146}$Sm, and $^{244}$Pu; Clayton, 1985, 1988; Nittler and Dauphas, 2006; Huss et al., 2009) or may have been produced by particle irradiation around the young Sun (*e.g.* $^{10}$Be and $^{36}$Cl, McKeegan et al., 2000; Hsu et al., 2006; Jacobsen et al., 2011). Others such as $^{26}$Al, which abundances exceed expetations for inheritance from galactic background (Meyer and Clayton, 2000; Wasserburg et al., 2006; Huss et al., 2009) and production by particle irradiation (Marhas et al., 2002; Duprat and Tatischeff 2007), were most likely injected by a nearby stellar source.

Iron-60 ($t_{1/2}$=2.62 Myr; Rugel et al., 2009) has been the focus of much work recently because it cannot be produced by particle irradiation in the solar protoplanetary disk (Lee et al., 1998) and a high $^{60}$Fe/$^{56}$Fe ratio in meteorites (>3×10$^{-7}$) would tie the birth of the solar system to the explosion of a nearby supernova (Wasserburg et al., 1998). However, the initial $^{60}$Fe/$^{56}$Fe ratio at the time of solar system formation is highly uncertain with estimates that span over two orders of magnitude (Birck and Lugmair,

1988; Shukolyukov and Lugmair, 1993a, b; Tachibana et al., 2003, 2006; Mostefaoui et al., 2004; 2005; Sugiura et al., 2006; Quitté et al. 2007, 2010, 2011; Guan et al., 2007; Tang and Dauphas, 2011a, b, 2012; Spivak-Birndorf et al., 2011, 2012; Moynier et al., 2011; Marhas and Mishra, 2012; Mishra and Chaussidon, 2012; Telus et al., 2012). Time zero in early solar system chronology is marked by calcium-aluminum-rich inclusions (CAIs), which are the oldest solar system solids. In CAIs, the only significant carriers of Fe and Ni are opaque assemblages (previously known as Fremdlinge; Palme and Wlotzka 1976; Wark and Lovering 1976; El Goresy et al., 1978; Blum et al., 1988; Sylvester et al., 1990) and smaller sub-micron refractory metal nuggets (Blander et al., 1980; Wark 1986; Berg et al., 2009), which have low Fe/Ni ratios and are therefore not very well suited to constrain the $^{60}$Fe/$^{56}$Fe ratio. Another complication is the fact that CAIs display isotopic anomalies for $^{62}$Ni that correlate with $^{60}$Ni, so it is difficult to ascribe $^{60}$Ni isotope variations solely to decay of $^{60}$Fe (Birck and Lugmair, 1988; Quitté et al., 2007). For these reasons, Birck and Lugmair (1988) and Quitté et al. (2007) were only able to establish an upper-limit on the initial $^{60}$Fe/$^{56}$Fe ratio of less than ~1.6 ×10$^{-6}$.

The first evidence for $^{60}$Fe in solar system material was found by thermal ionization mass spectrometry (TIMS) in the form of excess $^{60}$Ni in mineral separates from eucrites (Shukolyukov and Lugmair, 1993a, b), a group of basaltic meteorites that are thought to come from the crust of asteroid Vesta (McCord et al., 1970; De Sanctis et al., 2012; Russell et al., 2012). Internal isochrons in the eucrites Chervony Kut and Juvinas yielded initial $^{60}$Fe/$^{56}$Fe ratios of 3.9×10$^{-9}$ and 4.3×10$^{-10}$, respectively. However, the closure time for the $^{60}$Fe-$^{60}$Ni system in these meteorites was unknown, hindering a reliable estimate of the initial $^{60}$Fe/$^{56}$Fe ratio at solar system birth (Mostefaoui et al., 2004). Subsequent studies have yielded widely variable initial $^{60}$Fe/$^{56}$Fe ratio estimates.

*In situ* measurements by secondary ion mass spectrometry (SIMS) of constituents of unequilibrated ordinary chondrites, such as pyroxenes in chondrules, have given elevated $^{60}$Fe/$^{56}$Fe initial ratios at CAI formation of around ~6×10$^{-7}$ (Tachibana et al., 2003, 2006; Mostefaoui et al., 2004, 2005; Guan et al., 2007; Marhas and Mishra, 2012; Mishra et al., 2010; Mishra and Chaussidon, 2012). Ogliore et al. (2011) identified a statistical bias that affected some SIMS measurements (Tachibana et al., 2003, 2006), leading Telus et al. (2012) to revise the initial $^{60}$Fe/$^{56}$Fe ratio to 5×10$^{-8}$ or higher. More recent studies have also addressed these analytical issues and have found a high initial $^{60}$Fe/$^{56}$Fe ratio of ~6×10$^{-7}$ at solar system formation (Marhas and Mishra, 2012; Mishra and Chaussidon, 2012).

Measurements of differentiated achondrites (meteorites that were once molten) by various methods have given much lower ratios of ~2×10$^{-8}$ (Sugiura et al., 2006; Quitté et al., 2010, 2011; Tang and Dauphas, 2011a, 2011b, 2012; Spivak-Birndorf et al., 2011, 2012). The reason for this discrepancy is unknown but some have argued that this could be due to a heterogeneous distribution of $^{60}$Fe (Sugiura et al., 2006; Quitté et al., 2010). As discussed by Wasserburg et al. (1998), solving the issue of the initial abundance of $^{60}$Fe and of its distribution in early solar system materials is critical to assess the plausibility of the scenario of supernova-triggered solar system formation (Cameron and Truran, 1977; Boss and Keiser, 2012).

To study the initial abundance of $^{60}$Fe and its distribution in the disk, we have measured the Ni isotopic compositions of bulk HEDs, angrites and mineral separates from quenched angrite D'Orbigny. Motivated by the large discrepancy between achondrite and chondrite measurements, we have also studied the Ni isotopic compositions of whole rocks, chondrules and mineral separates from CBa (Bencubbin-type) chondrite Gujba, as well as two ordinary chondrites Semarkona (LL group) and NWA 5717 (ungrouped), which have experienced minimal thermal metamorphism (petrologic types 3.0 and 3.05, respectively, on a scale that starts at 3.0 and extends to 6; Grossman and Brearley 2005). Our results demonstrate that the $^{60}Fe/^{56}Fe$ initial ratio in the solar protoplanetary disk was $(11.5\pm2.6)\times10^{-9}$ and that $^{60}Fe$ was homogenously distributed among large planetary objects.

## 2. Methodology

Details of the methodology are available in Appendix A.

### 2.1 Sample preparation and digestion

Bulk achondrites (8 angrites, 9 HEDs, 2 ureillites, and 1 aubrite) weighing up to 1 g were first polished with abrasive paper and rinsed with acetone to get rid of fusion crust and other surface contaminants. D'Orbigny and NWA 5717 were fragmented by high voltage pulse power fragmentation (SelFrag) for mineral separation. The fragmented samples were processed with a hand magnet, sieves and sodium polytungstate solution. Using these procedures, metal, multiple silicate grain sizes and samples with different densities (below or above 3.10 g/cm$^3$) were separated. Bulk chondrules from Semarkona and NWA 5717 were handpicked for analyses. Six NWA 5717 entire chondrules and three silicate size fractions (100-166 μm, 166-200 μm and >200 μm) were washed before powdering in 1 M HCl for 30 minutes in order to remove surface-sited metal, sulfide, and other soluble phases. Twenty-one chondrules from CBa chondrite Gujba (a meteorite fall with little evidence of terrestrial alteration) were sampled using a New Wave Research Micromill apparatus. Silicate chondrules were crushed and the fragments with adhering metal sensitive to a hand magnet were removed before digestion. Whole rocks of several chondrites (1 CV, 1 CM, 1 EL, 1 EH, and 1 LL) and iron meteorites (1 IIAB, 1 IIIAB, 1 IVA, and 2 IVB) were also analyzed.

Iron meteorites and metal chondrules from Gujba were digested in aqua regia. All other samples were crushed in an agate mortar, and the sample powder was dissolved in a 5-30 mL mixture of concentrated HF-HNO$_3$ (in a 2:1 ratio by volume) in a Teflon beaker placed on a hot plate at ~90 °C for 5-10 days. The solution was subsequently evaporated to dryness and redissolved in a 5-30 mL mixture of concentrated HCl-HNO$_3$ (2:1 ratio). The solutions were dried down and taken back to solution with a minimum amount of concentrated HCl (~11 M) for loading on the first column.

### 2.2 Protocols for Ni and Fe purification

Chemical purification of Ni was achieved in a three-step procedure.

*(a) U/TEVA Cartridge*. The 2 mL U/Teva cartridges (length = 2.7 cm, diameter = 0.8 cm) were conditioned with 10 mL of concentrated HCl (~11 M). Approximately 3/4 of the sample solution in concentrated HCl was loaded onto the cartridge; the rest was kept for Fe/Ni determination, other analyses, and as a safety aliquot. The load solution was collected in clean Teflon beakers and an additional 10 mL of concentrated HCl was passed through the resin and collected in the same beaker. This eluate contained Ni together with other matrix elements (Horwitz et al., 1992).

*(b) Cation exchange.* The Ni elution cut from the first column chemistry was evaporated and re-dissolved in 5 mL of a mixture of 20 % 10 M HCl - 80% acetone (by volume) and loaded onto 5 ml pre-cleaned Bio-Rad AG50-X12 200-400 mesh hydrogen-form resin in teflon columns (length = 40 cm, diameter = 0.4 cm), previously conditioned with 10 mL 20 % 10 M HCl − 80 % acetone. After loading the sample solution and rinsing with 30 mL 20 % 10 M HCl − 80% acetone mixture, Ni was collected by eluting 150 mL of the HCl-acetone mixture (Strelow et al., 1971). This HCl-acetone column was repeated five times to ensure thorough separation of Mg from Ni.

*(c) Anion exchange.* Zinc was removed using a third column filled with 1 mL Bio-Rad AG1W-X8 anionic ion exchange resin (length = 2 cm, diameter = 0.8 cm) in 8 M HBr medium (Moynier et al., 2006).

Nickel blanks for the entire procedure ranged from 20 to 35 ng, which is insignificant relative to the amounts of Ni in the samples. Nickel yields for each step were close to 100 % and the overall yield of the procedure was 85 - 100 % except for Ibitira (around 60 %; Table S1). Achondrites and chondrules have very low Ni contents (*e.g.*, 0.5 μg for some HED meteorites) and the isotopic abundance of $^{64}$Ni could not be reliably measured in these samples but we were able to measure $^{64}$Ni in bulk chondrites and iron meteorites (also see Dauphas et al., 2008; Steele et al., 2011, 2012).

Tang et al. (2009) described the Fe purification protocol for measurement of isotopic anomalies. It uses a 10.5 cm long Teflon column (diameter = 0.62 cm) filled with 3 mL of AG1-X8 resin. Iron is retained on the column in >8 M HCl. Approximately 30 resin volumes (30 mL) of 4 M HCl is then passed through the column to eliminate Cu. Iron is finally eluted with 8 mL of 0.4 M HCl. The column separation procedure is repeated three times.

Iron isotopic fractionation measurements were done following the protocol of Dauphas et al. (2009).

## 2.3 Mass Spectrometry

The isotopic compositions of Ni and Fe were measured using a Neptune MC-ICPMS at the Origins Laboratory of the University of Chicago. L4, L2, L1, Ax, H1, H2 and H4 collectors were used to measure $^{57}$Fe$^+$, $^{58}$Ni$^+$+$^{58}$Fe$^+$, $^{60}$Ni$^+$, $^{61}$Ni$^+$, $^{62}$Ni$^+$, $^{64}$Ni$^+$+$^{64}$Zn$^+$ and $^{66}$Zn$^+$ ions, respectively. An amplifier with $10^{10}$ Ω resistor was used to measure ion intensities at mass 58. All other masses were collected using $10^{11}$ Ω amplifiers. Measurements were performed in medium resolution (MR) to resolve from argide interferences the $^{57}$Fe$^+$ flat-topped shoulder used to monitor and correct $^{58}$Fe interference on $^{58}$Ni. Samples in 0.3 M HNO$_3$ were introduced into the mass spectrometer with Ar + N$_2$ using an Aridus II inlet system at an uptake rate of ~100 μL/min. Jet sampler cones and X skimmer cones were used. The instrument sensitivity in these conditions for $^{58}$Ni was 70 V/ppm. One analysis consisted of 25 cycles, each acquisition lasting for 8.4 s. During a session, each sample solution was measured 6 to 17 times bracketed by SRM 986. The blank on peak zero was measured once during a session on the same acid used to dilute the samples and its intensity was subtracted from all standard and sample measurements. Internal normalization was used to correct mass-dependent isotopic fractionation by fixing $^{61}$Ni/$^{58}$Ni to 0.016730 or $^{62}$Ni/$^{58}$Ni to 0.053389 (Gramlich et al., 1989) using the exponential law (Maréchal et al., 1999).

The methodology for high precision measurements of Fe isotopic anomalies has been described previously (Dauphas et al., 2008). Faraday cups L4, L2, Ax, H1, H2 and H4 are set to collect $^{53}$Cr$^+$, $^{54}$Cr$^+$+$^{54}$Fe$^+$, $^{56}$Fe$^+$, $^{57}$Fe$^+$, $^{58}$Fe$^+$+$^{58}$Ni$^+$ and $^{60}$Ni$^+$, respectively. Ax is connected to a $10^{10}$ Ω amplifier to prevent saturation on $^{56}$Fe$^+$ and get relatively high intensities for low abundance isotopes. Other cups are connected to $10^{11}$ Ω amplifiers. Iron solutions (3 ppm) are introduced into the mass spectrometer in 0.3 M HNO$_3$ using an Aridus II inlet system. The samples, bracketed by IRMM-014 solution, are measured in sequences of 25 cycles of 8.4 s each. Each sample solution is measured 12-17 times in a sequence standard-sample-standard.

Approximately 20-25 % of the original sample solutions were kept for Fe/Ni ratio measurements by MC-ICPMS using both the bracketing and standard addition techniques. For Gujba, a fraction of that aliquot was also used for determination of iron isotopic fractionation (Dauphas et al., 2009).

## 3. Results

Table 1 compiles the Ni isotopic compositions, Fe/Ni ratios, and Fe isotopic fractionations measured in meteorites and terrestrial rock standards. Geostandards passed through the same column chemistry as meteoritic samples have normal Ni isotopic ratios, attesting to the accuracy of the measurements. Two groups of achondrites, ureilite (Kenna and EET83309) and aubrite (Bishopville), have low Fe/Ni ratios and did not yield any detectable $^{60}$Ni excess. Bulk chondrites and iron meteorites show small yet resolvable Ni isotopic anomalies of nucleosynthetic origin, in agreement with previous results (Dauphas et al., 2008; Regelous et al., 2008; Steele et al., 2011, 2012).

The intercepts and slopes of all $\varepsilon^{60}$Ni vs. $^{56}$Fe/$^{58}$Ni correlations were calculated using Isoplot (Ludwig 2012). Such correlations are used to estimate the initial $^{60}$Fe/$^{56}$Fe ratios in the samples using a relationship for decay of extinct radionuclides in closed systems,

$$\varepsilon^{60}Ni_{present} = \varepsilon^{60}Ni_0 + 2.596 \times 10^4 \left(\frac{^{60}Fe}{^{56}Fe}\right)_0 \left(\frac{^{56}Fe}{^{58}Ni}\right)_{present} \quad (1)$$

When the scatter around the regression was entirely explained by analytical uncertainties (Mean Square Weighted Deviation, MSWD~1), the uncertainties on the slope and the intercept of the isochron were calculated using a simple $1/\sigma^2$-weighted regression. When the scatter exceeded analytical uncertainties, model 3 of McIntyre et al. (1966) was used.

Nine bulk eucrites (Pasamonte, Juvinas, Camel Donga, Stennern, Ibitira, and Béréba) and diogenites (Shalka, Johnstown, and Tatahouine) were measured and the results are shown in Fig. 1A. Fe/Ni ratios range from 2,000 to 70,000, similar to what was documented previously for eucrites and diogenites (Wolf et al., 1983; Barrat et al., 2000; Shukolyukov and Lugmair, 1993). Quitté et al. (2011) had found a range of initial $^{60}$Fe/$^{56}$Fe ratios between $5\times10^{-10}$ and $5\times10^{-9}$ in bulk non-cumulate eucrites. Instead, in the present study (also see Tang and Dauphas 2011a, 2011b), the $^{60}$Ni excesses in bulk eucrites and diogenites correlate with $^{56}$Fe/$^{58}$Ni ratios and define a single errorchron of slope $^{60}$Fe/$^{56}$Fe = $(3.06\pm0.42)\times10^{-9}$ and intercept $\varepsilon^{60}$Ni = -0.10±0.14. The data points show some scatter around the best-fit line that cannot be completely explained by analytical uncertainties (MSWD=3.2). Two of the eucrites analyzed in this study (Pasamonte and Ibitira) have anomalous oxygen isotopic compositions, indicating that they came from different parent-bodies or that isotopic heterogeneity was preserved during magmatic differentiation on the HED parent body (Wiechert et al., 2004; Scott et al., 2009). Removing these two samples from the regression, the MSWD remains relatively high at 3.7 and the calculated $^{60}$Fe/$^{56}$Fe ratio does not change significantly ($3.06\times10^{-9}$ originally vs. $3.20\times10^{-9}$ after removal), so these two meteorites cannot be the source of the observed scatter. The Johnstown brecciated diogenite is rich in metal and highly siderophile elements, which was interpreted to reflect contamination by a chondritic impactor (Floran et al., 1981; Barrat et al., 2008; Dale et al., 2012). Its low Fe/Ni ratio is also consistent with this view. If we remove this sample from the regression, we obtain an isochron with an intercept of $\varepsilon^{60}Ni = -0.10 \pm 0.13$, a slope of $(3.45 \pm 0.32) \times 10^{-9}$, and a MSWD of 1.5. The cause for the scatter in the bulk HED regression is therefore due to the inclusion of a sample contaminated by a late addition of chondritic material. In the following, we exclude Johnstown from consideration in the bulk HED isochron.

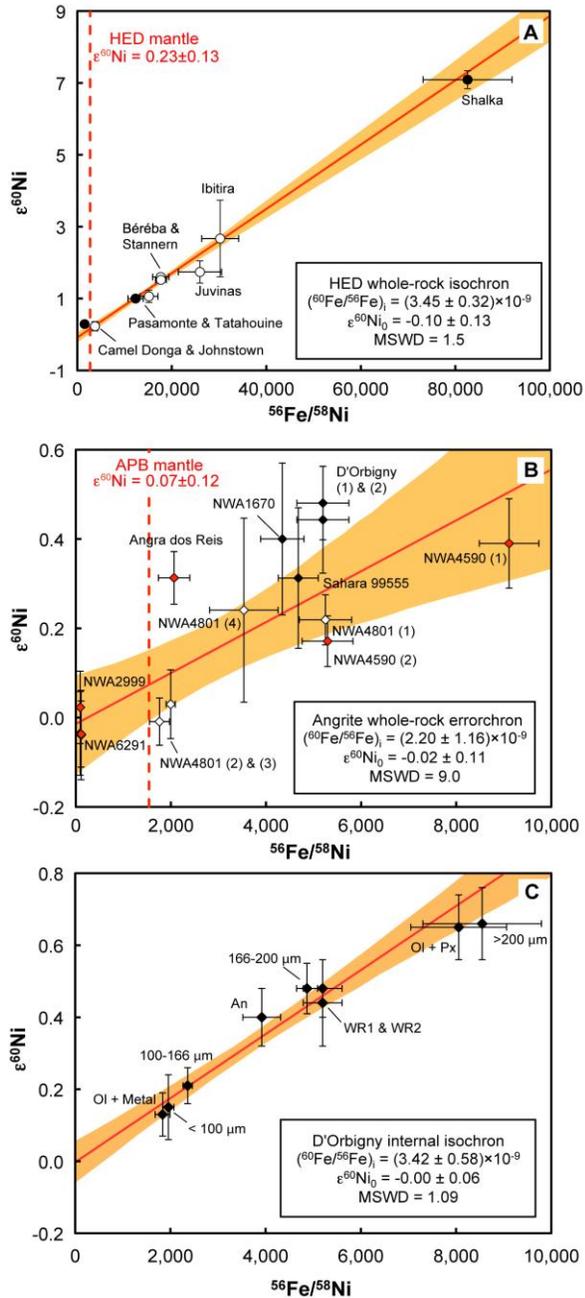

**Fig. 1.** $^{60}Fe$-$^{60}Ni$ isochron diagrams of achondrites. Ni isotopic ratios are reported in the $\varepsilon$-notation; $\varepsilon^{60}Ni=[(^{60}Ni/^{58}Ni)_{sample}/(^{60}Ni/^{58}Ni)_{standard}-1]\times10^4$, where $^{60}Ni/^{58}Ni$ ratios have been corrected for natural and laboratory-introduced mass fractionation by internal normalization to a constant $^{61}Ni/^{58}Ni$ ratio. The error bars represent 95 % confidence intervals. In $\varepsilon^{60}Ni$ vs. $^{56}Fe/^{58}Ni$ isochron diagrams, the intercept gives the initial Ni isotopic composition $\varepsilon^{60}Ni_0$ while the slope is proportional to the initial $^{60}Fe/^{56}Fe$ ratio; slope=$25,961\times(^{60}Fe/^{56}Fe)_0$. Live $^{60}Fe$ was detected in (A) bulk HED meteorites (open circles: eucrites, filled circles: diogenites; Johnstown is not included in the regression, see text for details), (B) bulk angrites [black filled diamonds: quenched angrites; red filled diamonds: plutonic angrites except NWA4801; open diamonds: different fragments of NWA4801; NWA4801 (1) and (4), olivine-pyroxene-rich fragments; (2) and (3), anorthite-rich fragments], (C) mineral separates of D'Orbigny (Ol+Metal, olivine with metal inclusions; An, anorthite; Ol, olivine; Px, pyroxene; WR, whole rock, the other labels represent fractions with different grain sizes). Following correction for decay after CAI formation, these measurements give a $^{60}Fe/^{56}Fe$ ratio at the birth of the solar system of $(1.17\pm0.26)\times10^{-8}$ (considering only the D'Orbigny and bulk HED isochrons).

Table 1. (continued)

| Sample Name | Sample Mass (mg) | Fe/Ni (at.) | $^{56}Fe/^{58}Ni$ | $\varepsilon^{60}Ni$ | Norm. $^{61}Ni/^{58}Ni$ $\varepsilon^{62}Ni$ | $\varepsilon^{64}Ni$ | $\varepsilon^{60}Ni$ | Norm. $^{62}Ni/^{58}Ni$ $\varepsilon^{61}Ni$ | $\varepsilon^{64}Ni$ | n Replicates | $\delta^{56}Fe$ | $\delta Fe$ | $\delta^{57}Fe$ |
|---|---|---|---|---|---|---|---|---|---|---|---|---|---|
| **Angrites** | | | | | | | | | | | | | |
| Angra dos Reis | 188 | 1572 | 2066 ± 245 | 0.31 ± 0.06 | 0.02 ± 0.08 | | 0.30 ± 0.04 | -0.02 ± 0.06 | | 14 | | | |
| NWA 1670 | 88 | 3338 | 4343 ± 455 | 0.40 ± 0.17 | 0.05 ± 0.27 | | 0.38 ± 0.10 | -0.04 ± 0.20 | | 12 | | | |
| NWA 2999 (1) | 20 | 70 | 91.5 ± 9.0 | 0.02 ± 0.08 | 0.12 ± 0.22 | | -0.04 ± 0.10 | -0.09 ± 0.17 | | 14 | | | |
| NWA 2999 (2) | 20 | | | -0.04 ± 0.09 | -0.07 ± 0.20 | | 0.01 ± 0.10 | 0.08 ± 0.15 | | 9 | | | |
| NWA 4590 (1) | 510 | 7004 | 9114 ± 627 | 0.39 ± 0.10 | -0.01 ± 0.20 | | 0.40 ± 0.07 | 0.01 ± 0.15 | | 12 | | | |
| NWA 4590 (2) | 200 | 4070 | 5296 ± 535 | 0.17 ± 0.06 | 0.07 ± 0.14 | | 0.14 ± 0.06 | -0.05 ± 0.11 | | 10 | | | |
| NWA4801 (1) | 450 | 4035 | 5250 ± 551 | 0.22 ± 0.06 | 0.01 ± 0.08 | | 0.21 ± 0.05 | -0.01 ± 0.06 | | 12 | | | |
| NWA4801 (2) | 290 | 1350 | 1758 ± 210 | -0.01 ± 0.05 | -0.03 ± 0.11 | | 0.00 ± 0.05 | 0.02 ± 0.08 | | 12 | | | |
| NWA4801 (3) | 460 | 1533 | 1995 ± 97 | 0.03 ± 0.08 | 0.04 ± 0.15 | | 0.01 ± 0.07 | -0.03 ± 0.11 | | 12 | | | |
| NWA4801 (4) | 910 | 2716 | 3534 ± 524 | 0.24 ± 0.21 | -0.07 ± 0.73 | | 0.27 ± 0.36 | 0.05 ± 0.56 | | 6 | | | |
| NWA 6291 (1) | 17 | 85 | 110.5 ± 9.9 | -0.04 ± 0.09 | 0.06 ± 0.16 | | -0.08 ± 0.09 | -0.06 ± 0.14 | | 13 | | | |
| NWA 6291 (2) | 17 | | | -0.04 ± 0.07 | -0.20 ± 0.32 | | 0.01 ± 0.06 | 0.07 ± 0.16 | | 9 | | | |
| SAH 99555 | 368 | 3376 | 4680 ± 311 | 0.31 ± 0.16 | 0.07 ± 0.22 | | 0.28 ± 0.16 | -0.05 ± 0.17 | | 6 | | | |
| D'Orbigny | | | | | | | | | | | | | |
| WR1 | 285 | 3994 | 5197 ± 406 | 0.48 ± 0.08 | -0.02 ± 0.20 | | 0.49 ± 0.11 | 0.02 ± 0.07 | | 8 | | | |
| WR2 | 285 | | | 0.44 ± 0.12 | 0.17 ± 0.42 | | 0.35 ± 0.26 | -0.13 ± 0.32 | | 8 | | | |
| <100 μm | 165 | 1504 | 1957 ± 119 | 0.15 ± 0.09 | 0.00 ± 0.17 | | 0.15 ± 0.06 | 0.00 ± 0.13 | | 14 | | | |
| 100-166 μm | 195 | 1816 | 2362 ± 97 | 0.21 ± 0.05 | 0.08 ± 0.11 | | 0.17 ± 0.04 | -0.06 ± 0.08 | | 14 | | | |
| 166-200 μm | 127 | 3742 | 4869 ± 221 | 0.48 ± 0.07 | -0.05 ± 0.12 | | 0.50 ± 0.06 | 0.04 ± 0.09 | | 14 | | | |
| >200 μm | 225 | 6569 | 8548 ± 1245 | 0.66 ± 0.10 | 0.12 ± 0.17 | | 0.60 ± 0.09 | -0.09 ± 0.13 | | 14 | | | |
| Anorthite | 338 | 3011 | 3919 ± 398 | 0.40 ± 0.08 | 0.08 ± 0.14 | | 0.35 ± 0.04 | -0.06 ± 0.10 | | 15 | | | |
| Metal* | 5.5 | 1410 | 1835 ± 156 | 0.13 ± 0.06 | 0.10 ± 0.11 | | 0.08 ± 0.03 | -0.08 ± 0.08 | | 14 | | | |
| Olivine + Pyroxene | 308 | 6191 | 8056 ± 1006 | 0.65 ± 0.09 | 0.08 ± 0.17 | | 0.61 ± 0.04 | -0.06 ± 0.13 | | 15 | | | |
| **Iron Meteorites** | | | | | | | | | | | | | |
| Coahuila (IIAB) | 44 | 17 | 24 | -0.04 ± 0.04 | -0.06 ± 0.09 | -0.21 ± 0.17 | -0.01 ± 0.02 | 0.04 ± 0.07 | -0.13 ± 0.08 | 17 | | | |
| Henbury (IIIAB) | 12 | 12 | 17 | -0.08 ± 0.04 | -0.15 ± 0.09 | -0.33 ± 0.14 | -0.06 ± 0.02 | -0.09 ± 0.04 | -0.11 ± 0.10 | 17 | | | |
| Gibeon (IVA) | 27 | 12 | 17 | -0.06 ± 0.04 | -0.08 ± 0.07 | -0.19 ± 0.14 | -0.02 ± 0.02 | 0.06 ± 0.05 | -0.06 ± 0.08 | 17 | | | |
| Cape of Good Hope (IVB) | 16 | 5 | 7 | -0.13 ± 0.03 | 0.07 ± 0.06 | 0.21 ± 0.12 | -0.16 ± 0.02 | -0.05 ± 0.05 | 0.11 ± 0.06 | 17 | | | |
| Tlacotepec (IVB) | 13 | 5 | 7 | -0.17 ± 0.04 | -0.00 ± 0.09 | 0.17 ± 0.13 | -0.17 ± 0.02 | 0.00 ± 0.07 | 0.17 ± 0.07 | 17 | | | |

Table 1. (continued)

| Sample Name | Sample Mass (mg) | Fe/Ni (at.) | $^{56}Fe/^{58}Ni$ | $\varepsilon^{60}Ni$ | Norm. $^{61}Ni/^{58}Ni$ $\varepsilon^{62}Ni$ | $\varepsilon^{64}Ni$ | $\varepsilon^{60}Ni$ | Norm. $^{62}Ni/^{58}Ni$ $\varepsilon^{61}Ni$ | $\varepsilon^{64}Ni$ | $n$ Replicates | $\delta^{56}Fe$ | $\delta Fe$ | $\delta^{57}Fe$ |
|---|---|---|---|---|---|---|---|---|---|---|---|---|---|
| **Bulk Chondrites** | | | | | | | | | | | | | |
| Allende (CV3) | 114 | 19 | 25 | $-0.14 \pm 0.03$ | $0.11 \pm 0.07$ | $0.30 \pm 0.10$ | $-0.20 \pm 0.02$ | $-0.08 \pm 0.05$ | $0.15 \pm 0.08$ | 17 | | | |
| Murchison (CM2) | 90 | 19 | 25 | $-0.13 \pm 0.04$ | $0.09 \pm 0.08$ | $0.26 \pm 0.13$ | $-0.17 \pm 0.03$ | $-0.06 \pm 0.06$ | $0.13 \pm 0.08$ | 17 | | | |
| Khairpur (EL6) | 93 | 18 | 24 | $-0.05 \pm 0.05$ | $-0.04 \pm 0.09$ | $-0.01 \pm 0.16$ | $-0.03 \pm 0.02$ | $0.03 \pm 0.07$ | $0.06 \pm 0.11$ | 17 | | | |
| St Mark's (EH5) | 79 | 18 | 24 | $-0.02 \pm 0.05$ | $0.04 \pm 0.10$ | $0.10 \pm 0.14$ | $-0.04 \pm 0.03$ | $-0.03 \pm 0.07$ | $0.04 \pm 0.10$ | 17 | | | |
| Chainpur (LL3.4) | 40 | 19 | $26 \pm 4$ | $-0.06 \pm 0.03$ | $-0.03 \pm 0.08$ | | $-0.06 \pm 0.02$ | $0.02 \pm 0.06$ | | 11 | | | |
| **Semarkona chondrules (LL3.0)** | | | | | | | | | | | | | |
| Sc-10-2 | 0.3 | 38 | $50 \pm 2.9$ | $-0.03 \pm 0.12$ | $-0.11 \pm 0.17$ | | $0.03 \pm 0.12$ | $0.09 \pm 0.13$ | | 10 | | | |
| Sc-30-6 | 0.2 | 13 | $17 \pm 1.0$ | $-0.12 \pm 0.25$ | $-0.38 \pm 0.42$ | | $0.08 \pm 0.14$ | $0.28 \pm 0.32$ | | 10 | | | |
| **NWA 5717 (Ungrouped 3.05)** | | | | | | | | | | | | | |
| Chondrule 3 | 6.5 | 88 | $114 \pm 6.7$ | $-0.02 \pm 0.03$ | $0.02 \pm 0.06$ | | $-0.03 \pm 0.04$ | $-0.01 \pm 0.04$ | | 14 | | | |
| Chondrule 8 | 4.7 | 25 | $32 \pm 1.9$ | $-0.04 \pm 0.04$ | $-0.01 \pm 0.08$ | | $-0.03 \pm 0.04$ | $0.01 \pm 0.06$ | | 14 | | | |
| Chondrule 1w** | 20 | 23 | $30 \pm 3.2$ | $-0.04 \pm 0.04$ | $-0.01 \pm 0.09$ | | $-0.04 \pm 0.05$ | $0.00 \pm 0.08$ | | 13 | | | |
| Chondrule 2w | 12 | 16 | $21 \pm 1.9$ | $-0.02 \pm 0.04$ | $-0.03 \pm 0.14$ | | $-0.01 \pm 0.07$ | $0.02 \pm 0.11$ | | 13 | | | |
| Chondrule 3w | 10 | 59 | $77 \pm 7.4$ | $-0.06 \pm 0.06$ | $0.03 \pm 0.07$ | | $-0.04 \pm 0.05$ | $0.02 \pm 0.07$ | | 13 | | | |
| Chondrule 4w | 11 | 22 | $30 \pm 2.1$ | $-0.03 \pm 0.05$ | $0.01 \pm 0.10$ | | $-0.03 \pm 0.03$ | $-0.01 \pm 0.07$ | | 13 | | | |
| Chondrule 5w | 10 | 31 | $41 \pm 4.0$ | $-0.02 \pm 0.04$ | $0.04 \pm 0.10$ | | $-0.04 \pm 0.04$ | $-0.03 \pm 0.07$ | | 13 | | | |
| Chondrule 6w | 12 | 24 | $31 \pm 2.2$ | $-0.01 \pm 0.04$ | $0.04 \pm 0.10$ | | $-0.01 \pm 0.05$ | $0.00 \pm 0.09$ | | 13 | | | |
| Magnetic fraction | 9 | 14 | $18 \pm 0.8$ | $-0.07 \pm 0.04$ | $-0.03 \pm 0.08$ | | $-0.05 \pm 0.06$ | $0.02 \pm 0.06$ | | 18 | | | |
| Silicates <100 μm | 15.8 | 29 | $38 \pm 0.8$ | $-0.03 \pm 0.05$ | $0.04 \pm 0.10$ | | $-0.05 \pm 0.06$ | $-0.01 \pm 0.07$ | | 16 | | | |
| 100-166 μm | 28.8 | 41 | $54 \pm 5.0$ | $-0.03 \pm 0.04$ | $-0.07 \pm 0.07$ | | $0.00 \pm 0.05$ | $0.05 \pm 0.06$ | | 18 | | | |
| 166-200 μm | 66.5 | 46 | $59 \pm 4.0$ | $-0.02 \pm 0.08$ | $-0.08 \pm 0.12$ | | $0.04 \pm 0.06$ | $0.06 \pm 0.09$ | | 16 | | | |
| >200 μm | 35.3 | 40 | $52 \pm 3.6$ | $-0.03 \pm 0.04$ | $0.04 \pm 0.07$ | | $-0.05 \pm 0.06$ | $-0.03 \pm 0.07$ | | 14 | | | |
| 100-166 μm w | 16.2 | 54 | $70 \pm 6.2$ | $-0.04 \pm 0.04$ | $-0.05 \pm 0.12$ | | $-0.01 \pm 0.06$ | $0.04 \pm 0.09$ | | 15 | | | |
| 166-200 μm w | 50.3 | 56 | $73 \pm 5.1$ | $-0.03 \pm 0.04$ | $-0.04 \pm 0.08$ | | $-0.01 \pm 0.02$ | $0.03 \pm 0.07$ | | 15 | | | |
| >200 μm w | 28.3 | 51 | $66 \pm 6.0$ | $-0.02 \pm 0.05$ | $0.00 \pm 0.12$ | | $-0.02 \pm 0.02$ | $0.00 \pm 0.09$ | | 15 | | | |

Table 1. (continued)

| Sample Name | Sample Mass (mg) | Fe/Ni (at.) | $^{56}Fe/^{58}Ni$ | $\varepsilon^{60}Ni$ | Norm. $^{61}Ni/^{58}Ni$ $\varepsilon^{62}Ni$ | $\varepsilon^{64}Ni$ | $\varepsilon^{60}Ni$ | Norm. $^{62}Ni/^{58}Ni$ $\varepsilon^{61}Ni$ | $\varepsilon^{64}Ni$ | n Replicates | δFe $\delta^{56}Fe$ | $\delta^{57}Fe$ |
|---|---|---|---|---|---|---|---|---|---|---|---|---|
| **Gujba** | | | | | | | | | | | | |
| Metal (1) | 94.5 | 16 | 21 ± 2 | -0.20 ± 0.06 | 0.17 ± 0.13 | | -0.29 ± 0.05 | -0.13 ± 0.10 | | 12 | -0.130 ± 0.029 | -0.173 ± 0.036 |
| Metal (2) | 32.9 | 13 | 18 ± 2 | -0.19 ± 0.06 | 0.13 ± 0.11 | | -0.26 ± 0.05 | -0.10 ± 0.08 | | 12 | -0.056 ± 0.037 | -0.054 ± 0.050 |
| Metal (3) | 20.6 | 10 | 13 ± 2 | -0.18 ± 0.06 | 0.17 ± 0.14 | | -0.27 ± 0.03 | -0.13 ± 0.11 | | 15 | -0.170 ± 0.049 | -0.237 ± 0.068 |
| Chondrule 1 | 118.9 | 608 | 791 ± 45 | -0.11 ± 0.02 | 0.05 ± 0.11 | | -0.12 ± 0.02 | -0.03 ± 0.04 | | 12 | 0.209 ± 0.029 | 0.322 ± 0.036 |
| Chondrule 2 | 158.4 | 58 | 75 ± 8 | -0.11 ± 0.05 | 0.07 ± 0.13 | | -0.14 ± 0.04 | -0.05 ± 0.10 | | 12 | 0.149 ± 0.029 | 0.225 ± 0.036 |
| Chondrule 3 | 31.8 | 28 | 37 ± 5 | -0.14 ± 0.06 | 0.08 ± 0.08 | | -0.18 ± 0.04 | -0.06 ± 0.06 | | 12 | | |
| Chondrule 4 | 28.5 | 32 | 41 ± 5 | -0.15 ± 0.04 | -0.02 ± 0.10 | | -0.14 ± 0.03 | 0.02 ± 0.08 | | 12 | | |
| Chondrule 5 | 18.4 | 29 | 38 ± 3 | -0.10 ± 0.04 | -0.05 ± 0.09 | | -0.08 ± 0.04 | 0.04 ± 0.07 | | 12 | 0.181 ± 0.026 | 0.295 ± 0.033 |
| Chondrule 6 | 48.1 | 36 | 48 ± 4 | -0.17 ± 0.06 | 0.01 ± 0.12 | | -0.18 ± 0.03 | -0.01 ± 0.09 | | 12 | 0.276 ± 0.029 | 0.446 ± 0.036 |
| Chondrule 7 | 44 | 79 | 102 ± 5 | -0.10 ± 0.05 | -0.02 ± 0.13 | | -0.09 ± 0.04 | 0.02 ± 0.09 | | 12 | 0.320 ± 0.037 | 0.500 ± 0.050 |
| Chondrule 8 | 38 | 51 | 65 ± 4 | -0.12 ± 0.09 | 0.08 ± 0.18 | | -0.15 ± 0.05 | -0.05 ± 0.14 | | 12 | 0.144 ± 0.026 | 0.225 ± 0.033 |
| Chondrule 9 | 36.2 | 55 | 72 ± 5 | -0.11 ± 0.08 | 0.16 ± 0.16 | | -0.19 ± 0.04 | -0.12 ± 0.12 | | 12 | 0.112 ± 0.026 | 0.185 ± 0.033 |
| Chondrule GL | 183.9 | 300 | 391 ± 50 | -0.15 ± 0.05 | 0.07 ± 0.09 | | -0.18 ± 0.03 | -0.05 ± 0.06 | | 15 | 0.212 ± 0.031 | 0.334 ± 0.055 |
| Chondrule G2-1 | 121.9 | 37 | 49 ± 4 | -0.17 ± 0.07 | 0.08 ± 0.11 | | -0.21 ± 0.02 | -0.06 ± 0.08 | | 15 | 0.737 ± 0.031 | 1.120 ± 0.055 |
| Chondrule G2-2 | 89.6 | 173 | 226 ± 18 | -0.18 ± 0.04 | 0.04 ± 0.04 | | -0.20 ± 0.04 | -0.03 ± 0.03 | | 15 | 0.277 ± 0.031 | 0.428 ± 0.055 |
| Chondrule G4-1 | 131 | 43 | 57 ± 5 | -0.16 ± 0.04 | 0.11 ± 0.11 | | -0.22 ± 0.04 | -0.08 ± 0.08 | | 15 | 0.400 ± 0.031 | 0.605 ± 0.055 |
| Chondrule G4-2 | 121.7 | 23 | 30 ± 3 | -0.16 ± 0.03 | 0.07 ± 0.07 | | -0.19 ± 0.02 | -0.05 ± 0.05 | | 15 | 0.354 ± 0.031 | 0.428 ± 0.055 |
| Chondrule G4-3 | 159.5 | 107 | 139 ± 14 | -0.13 ± 0.04 | -0.01 ± 0.09 | | -0.13 ± 0.03 | -0.01 ± 0.08 | | 15 | 0.813 ± 0.031 | 1.121 ± 0.055 |
| Chondrule G4-4 | 25.2 | 29 | 38 ± 5 | -0.11 ± 0.03 | 0.09 ± 0.10 | | -0.16 ± 0.04 | -0.07 ± 0.08 | | 15 | 0.013 ± 0.049 | 0.040 ± 0.068 |
| Chondrule GS-1 | 113.8 | 34 | 46 ± 4 | -0.18 ± 0.04 | 0.04 ± 0.06 | | -0.20 ± 0.03 | -0.03 ± 0.05 | | 15 | 0.904 ± 0.047 | 1.370 ± 0.064 |
| Chondrule GS-2 | 106.4 | 32 | 41 ± 3 | -0.14 ± 0.05 | 0.09 ± 0.12 | | -0.18 ± 0.06 | -0.07 ± 0.09 | | 15 | 0.758 ± 0.049 | 1.155 ± 0.068 |

Note: $\varepsilon^i Ni = ([^iNi/^{58}Ni]_{sample}/[^iNi/^{58}Ni]_{SRM986}-1) \times 10^4$ ; $\delta^i Fe = ([^iFe/^{54}Fe]_{sample}/[^iFe/^{54}Fe]_{IRMM-014}-1) \times 10^3$. The uncertainties are 95% confidence intervals.

Metal* in D'Orbigny was separated by magnet and is composed of olivine with metal inclusions.

**Subscript w: Samples were washed with 1 M HCl for 30 minutes before digestion.

Eight bulk angrites (mafic igneous meteorites) were analyzed (quenched angrites NWA 1670, Sahara 99555, D'Orbigny and plutonic angrites NWA 2999, NWA 6291, NWA 4590, NWA 4801, Angra dos Reis), several of which show resolvable $^{60}$Ni excesses linearly correlated with $^{56}$Fe/$^{58}$Ni ratios (Fig. 1B). Data from plutonic angrites (Angra dos Reis, NWA 4590, NWA 4801) are very scattered (MSWD=9.0), reflecting a protracted magmatic history (*e.g.*, Nyquist et al., 2009; Kleine et al., 2012). With this caveat, we calculate a slope $^{60}$Fe/$^{56}$Fe ratio of $(2.20\pm1.16)\times10^{-9}$ and intercept $\varepsilon^{60}$Ni = -0.02±0.11 at the time of core formation/global silicate differentiation in the angrite parent-body. This is consistent with the measurements of three bulk angrites by Quitté et al. (2010) that gave an initial $^{60}$Fe/$^{56}$Fe ratio of $(3.1\pm0.8)\times10^{-9}$. Spivak-Birndorf et al. (2012) reported Ni isotope measurements of plutonic angrites NWA 2999, NWA 4801, and NWA 4590. Overall, their results agree with ours except for plutonic angrite NWA 4590, where no excess $^{60}$Ni was detected even in samples leached with HCl that have high Fe/Ni ratio. The reason for the discrepancy is unknown but could result from terrestrial weathering, differences in the lithologies measured or in sample handling procedures.

The $^{60}$Fe/$^{56}$Fe and initial $\varepsilon^{60}$Ni inferred from mineral separates from the D'Orbigny angrite are shown in Fig. 1C. This meteorite has a quenched texture indicative of rapid cooling (Mittlefehldt et al., 2002) and several chronometers point to early crystallization, which must have occurred 5.5±1.0 Myr after CAI (Glavin et al., 2004; Nyquist et al., 2009). The data points define an internal isochron (MSWD=1.09) of slope $^{60}$Fe/$^{56}$Fe = $(3.42\pm0.58)\times10^{-9}$ and intercept $\varepsilon^{60}$Ni = 0.00±0.06 at the time of closure to isotope exchange of the minerals investigated. This value agrees well with independent results reported for this meteorite, which give initial $^{60}$Fe/$^{56}$Fe ratios of $(4.1\pm2.6)\times10^{-9}$ (Quitté et al., 2010) and $(2.81\pm0.86)\times10^{-9}$ (Spivak-Birndorf et al., 2011).

Because Semarkona and NWA 5717 are weakly metamorphosed chondrites (LL3.0 and ungrouped ordinary chondrite 3.05), chondrules from these two meteorites are expected to show little disturbance in $^{60}$Fe-$^{60}$Ni systematics (Tachibana et al., 2006; Mishra et al., 2010; Marhas and Mishra, 2012; Mishra and Chaussidon, 2012). The analyzed chondrules were from a larger batch of dissolved chondrules and were selected based on their elevated Fe/Ni ratios (~12 to 80). No clearly resolvable correlation between $\varepsilon^{60}$Ni and Fe/Ni ratio was found in chondrules and mineral separates from unequilibrated ordinary chondrites (UOC), corresponding to a slope $^{60}$Fe/$^{56}$Fe ratio of $(1.04\pm1.31)\times10^{-8}$ and an intercept $\varepsilon^{60}$Ni value of -0.05±0.02 at chondrule formation (MSWD=0.66, Fig. 2; also see Tang and Dauphas 2011b). This is the first solid constraint on the initial $^{60}$Fe/$^{56}$Fe ratio in ordinary chondrites by another method than SIMS.

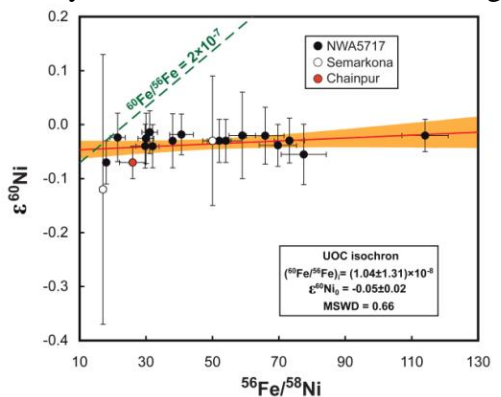

**Fig. 2.** $^{60}$Fe-$^{60}$Ni isochron diagrams of chondritic components from unequilibrated ordinary chondrites Semakona (LL3.0) and NWA 5717 (ungrouped ordinary chondrite 3.05). See Fig. 1 caption for notations. The inferred $^{60}$Fe/$^{56}$Fe ratio at UOC chondrule formation [$(1.04\pm1.31)\times10^{-8}$] is inconsistent with the values inferred previously from *in situ* measurements on the same specimens (~$2\times10^{-7}$ at chondrule formation, green dashed line) (Mostefaoui et al., 2004, 2005; Guan et al., 2007; Telus et al., 2012; Marhas and Mishra, 2012; Mishra and Chaussidon, 2012).

CB chondrites are thought to have formed 5.6±1.3 Myr after solar system formation ($^{53}$Mn-$^{53}$Cr age; Nyquist et al., 2009; Yamashita et al., 2010) from a vapor-melt plume produced by a large impact between asteroids (Kallemeyn et al., 2001; Campbell et al., 2002; Rubin et al., 2003; Krot et al., 2005). Despite the large spread in Fe/Ni ratios of Gujba chondrules (~16 to 587), no excess $^{60}$Ni was found corresponding to an initial $^{60}$Fe/$^{56}$Fe ratio of (2.05±2.31)×10$^{-9}$ at the time of CBa chondrule formation (errorchron in Fig. 3A). The scatter in $\varepsilon^{60}$Ni of Gujba chondrules cannot be entirely explained by analytical uncertainty (MSWD=2.9). Zipfel and Weyer (2007) measured the iron isotopic composition of silicate and metal in Gujba and found large isotopic fractionation between the two phases ($\delta^{56}$Fe values of ~+0.1 ‰ for silicate and ~-0.4 ‰ for metal), which they ascribed to kinetic isotope fractionation during condensation (Richter et al., 2009).

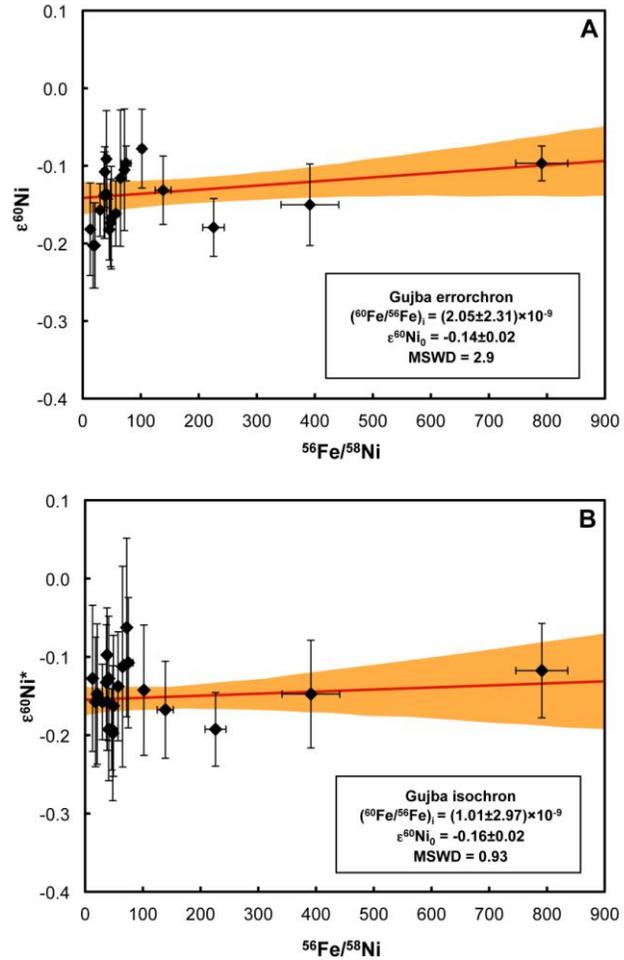

**Fig. 3**. $^{60}$Fe-$^{60}$Ni isochron diagrams of chondrules from CBa-chondrite Gujba. See Fig. 1 caption for notations. The top panel (A) shows the raw data, while the bottom panel (B) shows data corrected for mass-independent effects introduced by using an exponential mass fractionation law to correct condensation/evaporation- induced mass fractionation in Gujba metal and silicate, which would most likely a Rayleigh law with $\alpha = \sqrt{m_1/m_2}$. As discussed in the text and Appendix (Eq. 4 and B14), the correction applied is $\varepsilon^{60}$Ni* = $\varepsilon^{60}$Ni+0.5 ($\varepsilon^{62}$Ni+0.065). The slope of this isochron gives $^{60}$Fe/$^{56}$Fe = (1.01±2.97)×10$^{-9}$ at the time of formation of Gujba.

Alexander and Hewins (2004) measured the Ni and Fe isotopic fractionation in a zoned metal grain in Hammadah Al Hamrah 237, a CBb chondrite related to Gujba, and found large isotopic fractionations of Ni and Fe that were identical for both elements on a permil per amu basis (~8 ‰/amu range). Such fractionation would presumably follow a Rayleigh distillation with a fractionation factor that may scale as the square-root of the masses of the isotopes involved. The data are corrected for mass fractionation using an exponential law (generalized power-law with $n = 0$; Maréchal et al., 1999), which could introduce spurious isotope effects if the sample was affected by natural mass fractionation following a Rayleigh distillation with $\alpha = \sqrt{m_1/m_2}$ ($n = -0.5$; Appendix B, Eq. B13),

$$\varepsilon^{60}\text{Ni} \approx 5 \times (-0.5 - 0)\frac{(60-58)(60-61)}{58}F \approx 0.086F, \quad (2)$$

$$\varepsilon^{62}\text{Ni} \approx 5 \times (-0.5 - 0)\frac{(62-58)(62-61)}{58}F \approx -0.17F. \quad (3)$$

The degree of Ni isotopic fractionation was not measured. However, in CB chondrites this is well approximated by Fe (Alexander and Hewins 2004). Iron isotope measurements revealed large isotopic fractionation in Gujba chondrules and the relationship between $\varepsilon^{62}\text{Ni}$ and $F$ conforms to the expectation from Eq. 3 (Fig. 4A). A fractionation of ~0.5 ‰/amu on Ni isotopes would correspond to a 0.043 shift in $\varepsilon^{60}\text{Ni}$ (Eq. 2) and would be sufficient to introduce some scatter in the Gujba isochron. Equations 2 and B13 depend on assumptions regarding the fractionating species and mass fractionation laws. A more robust way to correct $\varepsilon^{60}\text{Ni}$ for inappropriate mass fractionation correction is to use $^{62}\text{Ni}$ as there should be a correlation between the shifts in $\varepsilon^{60}\text{Ni}$ and $\varepsilon^{62}\text{Ni}$ (divide Eq. 2 by Eq. 3; also see Appendix B, Eq. B14),

$$\varepsilon^{60}\text{Ni} \approx \frac{(60-58)(60-61)}{(62-58)(62-61)}\varepsilon^{62}\text{Ni} \approx -0.5\varepsilon^{62}\text{Ni}. \quad (4)$$

Measured $\varepsilon^{60}\text{Ni}$ and $\varepsilon^{62}\text{Ni}$ values conform to the expectation from this equation (Fig. 3B). The Ni isotopic composition was therefore corrected using the formula $\varepsilon^{60}\text{Ni}^* = \varepsilon^{60}\text{Ni} + 0.5\ (\varepsilon^{62}\text{Ni} - 0.065)$, where +0.065 is the $\varepsilon^{62}\text{Ni}$ value when $F = 0$ (Fig. 4A). The resulting isochron shows less scatter than the one using uncorrected values and the MSWD decreases from 2.9 to 0.93 (Fig. 3B). The slope and intercept give $^{60}\text{Fe}/^{56}\text{Fe} = (1.01 \pm 2.97) \times 10^{-9}$ and $\varepsilon^{60}\text{Ni}_0 = -0.16 \pm 0.02$. The results on Gujba are consistent with previous measurements but the range of Fe/Ni ratios is expanded to much higher values (up to 600; Table 1) than earlier work (Fe/Ni~15; Quitté et al., 2011) and as a result provide the first solid constraint on the initial $^{60}\text{Fe}/^{56}\text{Fe}$ ratio in the Gujba parent-body. Wielandt et al. (2012) reported an initial $^{60}\text{Fe}/^{56}\text{Fe}$ ratio of $(5.0 \pm 1.4) \times 10^{-8}$ in Gujba that is inconsistent with the results presented here.

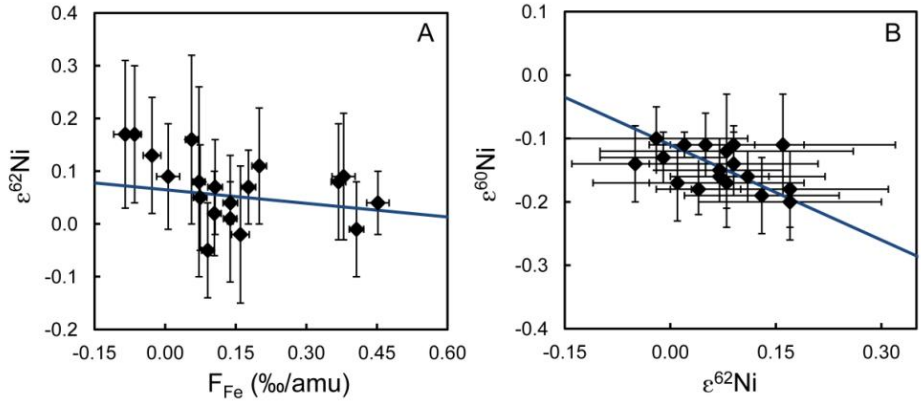

**Fig. 4.** Effect on mass fractionation laws on measured Ni isotopic anomalies in Gujba. (A) Correlation between $\varepsilon^{62}\text{Ni} = [(^{62}\text{Ni}/^{58}\text{Ni})_{sample}/(^{62}\text{Ni}/^{58}\text{Ni})_{standard} - 1] \times 10^4$ corrected for mass fractionation using an exponential law and the degree of mass fractionation for Fe isotopes $F_{Fe} = \delta^{56}\text{Fe}/2$, where $\delta^{56}\text{Fe} = [(^{56}\text{Fe}/^{54}\text{Fe})_{sample}/(^{56}\text{Fe}/^{54}\text{Fe})_{IRMM-014} - 1] \times 10^3$. (B) Correlation between $\varepsilon^{62}\text{Ni}$ and $\varepsilon^{60}\text{Ni}$. The blue lines correspond to the expected correlations given by Eqs. 3 and 4 (see text for details).

Table 2. Fe isotopic compositions of terrestrial standards, chondrites and achondrites.

| Sample Name | Sample Type | Normalized to $^{57}Fe/^{54}Fe$ | | $n$ Replicates |
|---|---|---|---|---|
| | | $\varepsilon^{56}Fe$ | $\varepsilon^{58}Fe$ | |
| **Terrestrial Standards** | | | | |
| BIR-1 | | -0.02 ± 0.19 | -0.11 ± 0.73 | 13 |
| DNC-1 | | -0.02 ± 0.11 | 0.02 ± 0.22 | 13 |
| WS-E | | -0.03 ± 0.30 | -0.37 ± 0.54 | 15 |
| **Angrites** | | | | |
| D'Orbigny | | -0.05 ± 0.21 | 0.45 ± 0.64 | 13 |
| NWA2999 | | 0.11 ± 0.20 | -0.08 ± 0.67 | 13 |
| NWA4801 | | -0.03 ± 0.30 | -0.24 ± 0.71 | 15 |
| NWA6291 | | -0.06 ± 0.17 | 0.08 ± 0.49 | 13 |
| SAH99555 | | 0.15 ± 0.27 | 0.23 ± 0.75 | 15 |
| **Eucrites** | | | | |
| Camel Donga | | 0.00 ± 0.11 | -0.11 ± 0.29 | 13 |
| Juvinas | | -0.08 ± 0.32 | 0.09 ± 0.56 | 15 |
| **Ureilites** | | | | |
| EET83309 | | -0.04 ± 0.07 | -0.01 ± 0.27 | 13 |
| Kenna | | -0.05 ± 0.11 | -0.07 ± 0.21 | 13 |
| **Chondrites** | | | | |
| Allende | CV3 | -0.03 ± 0.10 | -0.21 ± 0.43 | 15 |
| Chainpur | LL3.4 | -0.09 ± 0.08 | -0.18 ± 0.24 | 12 |
| Paragould | LL5 | -0.05 ± 0.15 | -0.25 ± 0.50 | 15 |
| Indarch | EH4 | 0.02 ± 0.14 | -0.04 ± 0.43 | 15 |
| Blithfield | EL6 | 0.02 ± 0.14 | -0.05 ± 0.19 | 13 |
| Yilmin | EL6 | -0.08 ± 0.12 | 0.13 ± 0.29 | 13 |
| **NWA5717** | Ungrouped 3.05 | | | |
| Chondrule 3 | | -0.07 ± 0.08 | -0.04 ± 0.26 | 13 |
| Chondrule 8 | | -0.08 ± 0.11 | -0.02 ± 0.17 | 13 |
| Magnetic fraction | | -0.10 ± 0.08 | -0.16 ± 0.18 | 12 |
| Silicates <100 µm | | 0.01 ± 0.08 | -0.06 ± 0.22 | 12 |
| 100-166 µm | | -0.02 ± 0.07 | 0.05 ± 0.22 | 12 |
| 166-200 µm | | -0.08 ± 0.08 | -0.17 ± 0.24 | 12 |
| >200 µm | | -0.01 ± 0.12 | -0.01 ± 0.23 | 13 |
| **Semarkona Chondrules** | LL3.0 | | | |
| Sc-10-2 | | 0.04 ± 0.11 | -0.06 ± 0.14 | 12 |
| Sc-30-6 | | -0.02 ± 0.15 | -0.04 ± 0.15 | 12 |
| **Allende Chondrules** | CV3 | | | |
| IH1 | | -0.10 ± 0.05 | 0.19 ± 0.20 | 15 |
| IH4 | | -0.05 ± 0.05 | 0.16 ± 0.17 | 15 |
| IH5 | | -0.07 ± 0.05 | -0.03 ± 0.20 | 15 |
| IH7 | | -0.00 ± 0.03 | -0.09 ± 0.12 | 15 |
| IH8 | | -0.03 ± 0.03 | -0.13 ± 0.16 | 15 |
| IH9 | | 0.00 ± 0.06 | 0.14 ± 0.28 | 15 |
| IH10 | | -0.15 ± 0.07 | 0.15 ± 0.17 | 15 |
| IH11 | | -0.04 ± 0.07 | 0.17 ± 0.29 | 15 |
| IH12 | | -0.13 ± 0.06 | 0.07 ± 0.18 | 15 |
| IH20 | | -0.16 ± 0.07 | 0.03 ± 0.22 | 15 |
| IH21 | | -0.07 ± 0.07 | 0.14 ± 0.19 | 15 |
| IH22 | | -0.06 ± 0.07 | 0.07 ± 0.17 | 15 |
| IH24 | | -0.04 ± 0.02 | -0.04 ± 0.14 | 15 |
| **Iron Meteorites** | | | | |
| Coahuila | IIIAB | -0.02 ± 0.03 | -0.17 ± 0.16 | 17 |
| Gibeon | IVA | 0.00 ± 0.23 | 0.12 ± 0.27 | 17 |
| Cape of Good Hope | IVB | -0.08 ± 0.04 | 0.01 ± 0.13 | 17 |
| Tlacotepec | IVB | -0.08 ± 0.05 | 0.13 ± 0.17 | 17 |

Note. $\varepsilon^i Fe$ is defined as $\varepsilon^i Fe = ([^i Fe/^{54}Fe]_{sample}/[^i Fe/^{54}Fe]_{IRMM-14} - 1) \times 10^3$. Dauphas et al. (2004, 2008) measured $\varepsilon^{56}Fe$ and $\varepsilon^{58}Fe$ in several meteorite groups (CI, CV, LL, Pallasites, IIAB and IIIAB iron meteorites)

In agreement with previous work (Dauphas et al., 2008), the iron isotopic compositions of all measured samples are identical to terrestrial composition after internal normalization (Table 2, Fig. 5). In particular, we did not detect any isotopic anomaly for the neutron-rich isotope $^{58}$Fe.

## 4. Discussion
### 4.1. A low and uniform initial $^{60}$Fe/$^{56}$Fe ratio in planetary bodies

A previous study had found a range of initial $^{60}$Fe/$^{56}$Fe ratios between $5 \times 10^{-10}$ and $5 \times 10^{-9}$ in bulk non-cumulate eucrites (Quitté et al., 2011). Instead, all our samples excluding Johnstown, which has been contaminated with chondritic material (Floran et al., 1981; Barrat et al., 2008; Dale et al., 2012), define a single isochron of slope $^{60}$Fe/$^{56}$Fe = $(3.45 \pm 0.32) \times 10^{-9}$ (Fig. 1A). The two eucrites that have anomalous oxygen isotopic compositions (Pasamonte and Ibitira; Wiechert et al., 2004; Scott et al., 2009) also plot on the same bulk rock isochron as other "normal" eucrites. This is consistent with the finding that Pasamonte and Ibitira also plot on the same bulk rock $^{53}$Mn-$^{53}$Cr ($t_{1/2}$=3.74 Myr; Honda and Imamura, 1971) isochron as other eucrites (Lugmair and Shukolyukov, 1998; Trinquier et al., 2008). If Pasamonte, Ibitira and "normal" HEDs were derived from different parent-bodies, this could indicate that mantle differentiation occurred simultaneously in these three objects. More likely, all HEDs are part of the same parent-body but the oxygen isotopic composition was never fully homogenized between mantle reservoirs.

The $^{60}$Fe-$^{60}$Ni bulk rock HED isochron records the time of global silicate differentiation in the mantle of Vesta, which is also marked by a well-defined isochron for the $^{53}$Mn-$^{53}$Cr short-lived radionuclide system (Lugmair and Shukolyukov, 1998; Trinquier et al., 2008). According to $^{53}$Mn-$^{53}$Cr systematics, mantle differentiation in the HED parent body took place $4.1 \pm 1.1$ Myr after CAI formation (Trinquier et al., 2008; Nyquist et al., 2009). Bulk rock isochrons are little susceptible to thermal disturbances, so one can safely use the time interval inferred from $^{53}$Mn-$^{53}$Cr systematics to back-calculate a $(^{60}\text{Fe}/^{56}\text{Fe})_0$ ratio at the time of CAI formation of $(1.02 \pm 0.32) \times 10^{-8}$. At such a low abundance, decay of $^{60}$Fe provided negligible heat to planetary objects (*i.e.*, <20 K increase in temperature) and $^{26}$Al ($t_{1/2}$=0.705±0.024 Myr) was the only radioactive heat source in the early solar system.

A $^{60}$Fe/$^{56}$Fe ratio of $(2.20 \pm 1.16) \times 10^{-9}$ at the time of core formation/global silicate differentiation in the angrite parent-body was obtained but data from plutonic angrites such as Angra dos Reis, NWA4801 and NWA4590 show significant scatter (Fig. 1B). Using the time interval between CAI and global silicate differentiation in the angrite parent-body from $^{53}$Mn-$^{53}$Cr systematics ($5.3 \pm 1.0$ Myr; Nyquist et al., 2009; Shukolyukov and Lugmair, 2007), an initial $(^{60}\text{Fe}/^{56}\text{Fe})_0$ ratio at the time of CAI formation of $(0.88 \pm 0.52) \times 10^{-8}$ is estimated. Although bulk angrites are not ideally suited to estimate the initial $^{60}$Fe/$^{56}$Fe ratio at the birth solar system, the inferred ratio is identical to that calculated from HED meteorites.

We have also measured the Ni isotopic composition in mineral separates from the D'Orbigny meteorite. This angrite has a quenched texture indicative of rapid cooling (Mittlefehldt et al., 2002; Keil 2012). Several chronometers ($^{26}$Al-$^{26}$Mg, $^{53}$Mn-$^{53}$Cr, $^{182}$Hf-$^{182}$W, and $^{207}$Pb/$^{206}$Pb) point to early crystallization, which according to $^{53}$Mn-$^{53}$Cr systematics must have occurred $5.5 \pm 1.0$ Myr after CAI (Glavin et al., 2004; Nyquist et al.,

2009). The data points define an internal isochron of slope $^{60}Fe/^{56}Fe = (3.42\pm0.58)\times10^{-9}$, from which we can back calculate a $(^{60}Fe/^{56}Fe)_0$ ratio at the time of CAI formation of $(1.47\pm0.46)\times10^{-8}$ (Fig. 1C). This value is identical to that obtained in bulk HEDs, demonstrating that the parent-bodies of angrites and HEDs formed from a reservoir with the same initial $(^{60}Fe/^{56}Fe)_0$ ratio at CAI formation.

The initial $^{60}Fe/^{56}Fe$ ratio obtained from achondrite measurements [$(1.17\pm0.26)\times10^{-8}$ using data from bulk HEDs and mineral separates of D'Orbigny; *i.e.*, excluding bulk angrites] is much lower than that inferred from *in situ* measurements of chondrite components (Mostefaoui et al., 2004, 2005; Guan et al., 2007; Telus et al., 2012; Marhas and Mishra, 2012; Mishra and Chaussidon, 2012). As discussed below, this discrepancy cannot be due to heterogeneous distribution of $^{60}Fe$ as our measurements of unequilibrated ordinary chondrites Semarkona (3.0) and NWA 5717 (3.05) give an initial $^{60}Fe/^{56}Fe$ ratio consistent with achondrite measurements.

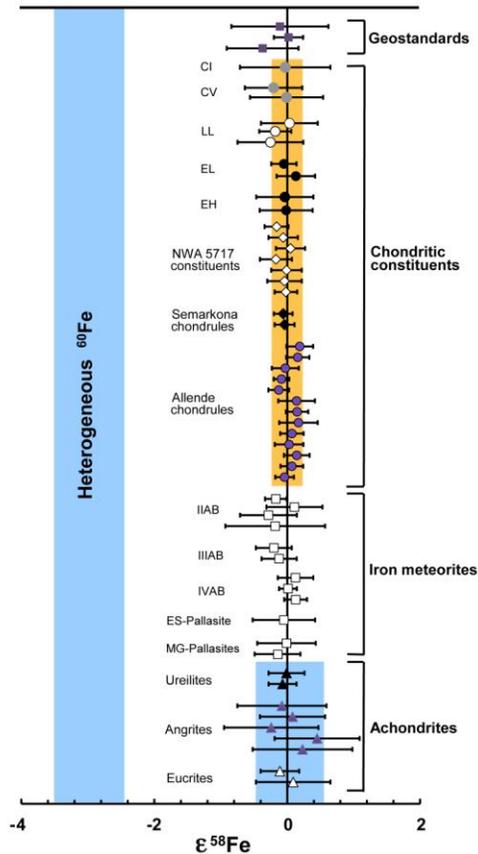

**Fig. 5.** $\varepsilon^{58}Fe$ in achondrites and chondritic samples (Table 2; Dauphas et al., 2004a, 2008). $\varepsilon^{58}Fe=[(^{58}Fe/^{54}Fe)_{sample}/(^{58}Fe/^{54}Fe)_{standard}-1]\times10^4$, where $^{58}Fe/^{54}Fe$ ratios have been corrected for mass fractionation by internal normalization to a constant $^{57}Fe/^{54}Fe$ ratio. Error bars are 95 % confidence intervals. If chondrites and achondrites had formed from different nebular reservoirs with distinct $^{60}Fe/^{56}Fe$ ratios of $6\times10^{-7}$ and $1\times10^{-8}$ respectively, then -3 $\varepsilon$ anomalies (blue bar labeled "Heterogeneous $^{60}Fe$") would be expected on $^{58}Fe$ in achondrites relative to chondrites (Dauphas et al., 2008). The fact these two meteorite groups have indistinguishable $\varepsilon^{58}Fe$ supports the view that $^{60}Fe$ was homogeneously distributed.

No clearly resolvable correlation between $\varepsilon^{60}Ni$ and Fe/Ni ratio was found in chondrules and mineral separates from UOC, corresponding to a $^{60}Fe/^{56}Fe$ ratio of $(1.04\pm1.31)\times10^{-8}$ at chondrule formation (Fig. 2). The timing of formation of these objects is not precisely known but most chondrules in ordinary chondrites were formed $2.4\pm2.0$ Myr after CAI (Kita et al., 2005; Yin et al., 2007; Nyquist et al., 2009; Villeneuve et al., 2009; Dauphas and Chaussidon, 2011). This translates into an initial $^{60}Fe/^{56}Fe$ ratio at CAI formation of $(1.96\pm2.69)\times10^{-8}$ in the region of the protoplanetary disk where ordinary chondrites formed.

CBa chondrites are thought to have formed $5.6\pm1.3$ Myr after solar system formation from a vapor-melt plume produced by a large impact between asteroids (Krot et al., 2005; Nyquist et al., 2009; Yamashita et al., 2010). Despite the large spread in Fe/Ni ratios of Gujba chondrules, no excess $^{60}Ni$ was found corresponding to an initial $^{60}Fe/^{56}Fe$ ratio of $(1.01\pm2.97)\times10^{-9}$ at the time of CBa chondrule formation (Fig. 3B). This constrains the $^{60}Fe/^{56}Fe$ ratio at CAI formation to $(0.45\pm1.32)\times10^{-8}$.

The values derived from chondrite measurements agree with the initial $^{60}$Fe/$^{56}$Fe ratio inferred from angrite and HED data and yield a weighted average of $(1.15\pm0.26) \times 10^{-8}$ (Table 3; because of protracted magmatic history, this value does not include the bulk angrite errorchron). The main source of uncertainty in the initial $^{60}$Fe/$^{56}$Fe is the initial $^{53}$Mn/$^{55}$Mn ratio used to back-calculate measured $^{60}$Fe/$^{56}$Fe ratios to the time of CAI formation (Fig. 6). The calculations above adopted an initial $^{53}$Mn/$^{55}$Mn ratio of $(9.1\pm1.7)\times10^{-6}$ (Nyquist et al., 2009). Others have proposed a lower initial $^{53}$Mn/$^{55}$Mn ratio of $(6.5\pm1.9)\times10^{-6}$ (Trinquier et al., 2008), which would give a lower $^{60}$Fe/$^{56}$Fe ratio of $(7.05\pm2.26)\times10^{-9}$. Regardless of this complication, our estimate of the initial $^{60}$Fe/$^{56}$Fe ratio at CAI formation is inconsistent with the high initial $^{60}$Fe/$^{56}$Fe ratio inferred from SIMS data (Mostefaoui et al., 2004, 2005; Guan et al., 2007; Telus et al., 2012; Marhas and Mishra, 2012; Mishra and Chaussidon, 2012). While the Fe/Ni ratios of the components studied here are much lower than those studied by *in situ* methods, our precision is superior and we should have detected excess $^{60}$Ni of +3.7 ε-unit in chondrites if the initial $^{60}$Fe/$^{56}$Fe ratio at CAI formation was $\sim 6\times10^{-7}$. Thus, our results indicate that different regions of the protoplanetary disk incorporated the same amounts of $^{60}$Fe. Variations in $^{60}$Fe abundance might be present at the subcentimeter scale as is suspected for example for $^{26}$Al in FUN CAIs, a type of refractory inclusion that contains Fractionated and Unknown Nuclear effects (MacPherson et al., 1995). Opaque assemblages in CAIs may allow us to constrain the abundance of $^{60}$Fe in CAIs but these have low Fe/Ni ratios (*e.g.*, Sylvester et al., 1990) and $^{60}$Fe-decay will only be detectable if $^{60}$Fe/$^{56}$Fe initial ratios are much higher than the value documented here.

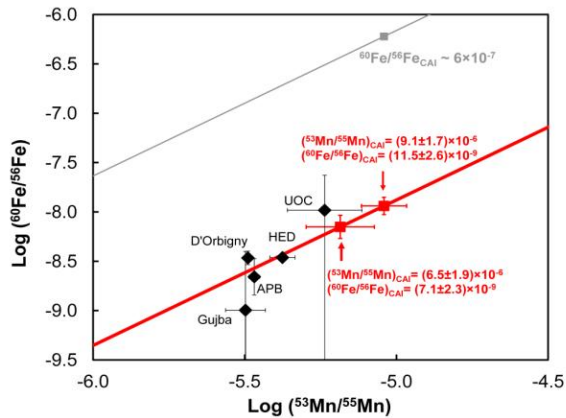

**Fig. 6.** Relationship between Log($^{60}$Fe/$^{56}$Fe) and Log($^{53}$Mn/$^{55}$Mn) initial ratios in the meteorite groups studied here (Table 3). The red line shows the expected correlation between these ratios for a system starting with homogeneous $^{60}$Fe/$^{56}$Fe and $^{53}$Mn/$^{55}$Mn ratios. The slope of this line is the ratio of the half-lives of $^{53}$Mn and $^{60}$Fe (=3.74/2.62=1.43). The two red dots are the estimated initial $^{60}$Fe/$^{56}$Fe ratios at CAI formation assuming $^{53}$Mn/$^{55}$Mn=$(9.1\pm1.7)\times10^{-6}$ (Nyquist et al., 2009) or $^{53}$Mn/$^{55}$Mn=$(6.5\pm1.9)\times10^{-6}$ (Trinquier et al., 2008). The grey line shows the expected evolution for a system starting at an initial ratio $^{60}$Fe/$^{56}$Fe of $\sim6\times10^{-7}$ (Mostefaoui et al., 2004, 2005; Guan et al., 2007; Marhas and Mishra, 2012; Mishra and Chaussidon, 2012). Our results exclude such a high ratio and demonstrate that $^{60}$Fe was homogeneous distributed in the meteorite groups investigated.

A further way to examine the question of the homogeneity of $^{60}$Fe in the early solar system is to measure the isotopic abundance of $^{58}$Fe (Dauphas et al., 2008). Indeed, in core-collapse supernovae, these two neutron-rich isotopes are produced by neutron-capture reactions on pre-existing Fe isotopes, therefore any heterogeneity in $^{60}$Fe should be accompanied by $^{58}$Fe isotope anomalies. If chondrites formed from a reservoir with $(^{60}$Fe/$^{56}$Fe$)_0\sim6\times10^{-7}$ while achondrites formed from a reservoir with $(^{60}$Fe/$^{56}$Fe$)_0\sim1\times10^{-8}$, the latter group of meteorites should show deficits in $^{58}$Fe of -3 ε-unit or more relative to chondrites (Dauphas et al., 2008). We have measured at high precision the isotopic ratio $^{58}$Fe/$^{54}$Fe in many meteorite groups, and have failed to detect any variation in this isotope (within ±0.5 ε-unit) (Fig. 5). Dauphas et al. (2010) speculated that the $^{60}$Fe/$^{56}$Fe ratio in the solar protoplanetary disk might be correlated with variations in $\varepsilon^{54}$Cr if $^{54}$Cr and $^{60}$Fe came from the same nearby star. Carbonaceous chondrites Allende (CV) and Orgueil (CI) have among the highest $\varepsilon^{54}$Cr values (+0.9 and +1.6, respectively; Trinquier et al., 2007) but have identical $\varepsilon^{58}$Fe values to achondrites (Fig. 5). This supports the view that achondrites and chondrites shared the same initial $^{60}$Fe/$^{56}$Fe ratio of $(1.15\pm0.26)\times10^{-8}$.

Table 3. Initial $^{60}$Fe/$^{56}$Fe ratio in the solar protoplanetary disk.

| Sample | $^{60}$Fe/$^{56}$Fe$_i$ from isochrons | $\Delta t_{Mn-Cr}$ (Myr) relative to CAI | Calculated $^{60}$Fe/$^{56}$Fe$_0$ at CAI formation* |
|---|---|---|---|
| Bulk HED meteorites | $(3.45\pm0.32)\times10^{-9}$ | $4.1\pm1.1$ | $(1.02\pm0.32)\times10^{-8}$ |
| Bulk angrites | $(2.20\pm1.16)\times10^{-9}$ | $5.3\pm1.0$ | $(0.88\pm0.52)\times10^{-8}$ |
| D'Orbigny minerals | $(3.42\pm0.58)\times10^{-9}$ | $5.5\pm1.0$ | $(1.47\pm0.46)\times10^{-8}$ |
| UOC chondrules & minerals | $(1.04\pm1.31)\times10^{-8}$ | $2.4\pm2.0$ | $(1.96\pm2.69)\times10^{-8}$ |
| Gujba chondrules | $(1.01\pm2.97)\times10^{-9}$ | $5.6\pm1.3$ | $(0.45\pm1.32)\times10^{-8}$ |
| Best estimate initial $^{60}$Fe/$^{56}$Fe at CAI formation**: | | | $(1.15\pm0.26)\times10^{-8}$ |

* $^{60}$Fe/$^{56}$Fe initial ratios are calculated using an initial $^{53}$Mn/$^{55}$Mn at CAI formation of $(9.1\pm1.7)\times10^{-6}$ (Nyquist et al., 2009). A lower $^{53}$Mn/$^{55}$Mn ratio of $(6.5\pm1.9)\times10^{-6}$ in the inner solar system (Trinquier et al., 2008) would yield $^{60}$Fe/$^{56}$Fe=$(7.05\pm2.26)\times10^{-9}$ at CAI formation.
** Calculated using initial $^{60}$Fe/$^{56}$Fe ratios at CAI formation excluding bulk angrites, which show disturbed $^{60}$Fe-$^{60}$Ni systematics.

**4.2. Chronology of core formation in the HED and angrite parent-bodies**

The $^{60}$Fe-$^{60}$Ni bulk HED isochron and bulk angrite errorchron record the $^{60}$Fe/$^{56}$Fe ratio at the time of mantle differentiation in their parent-bodies, the same events that are recorded by $^{53}$Mn-$^{53}$Cr systematics. Because metal-silicate partitioning can fractionate the Fe/Ni ratio, $^{60}$Fe-$^{60}$Ni systematics can also be used to establish the time of core formation in planetesimals. Let us consider a multi-stage model of core formation and mantle differentiation. From $t$=0 (CAI formation) to $t_{core}$, the parent-body evolves with chondritic composition (chondritic Fe/Ni ratio). At $t_{core}$, it differentiates into a mantle (high Fe/Ni ratio because Ni is more siderophile than Fe) and a core (low Fe/Ni ratio). The mantle then differentiates into distinct reservoirs leading to additional fractionation of Fe/Ni and Mn/Cr ratios, an event that is recorded by bulk rock $^{60}$Fe-$^{60}$Ni and $^{53}$Mn-$^{53}$Cr isochrons. A 2-stage model age of core formation can be calculated by building a two-point isochron between the bulk planetary object (assumed to be chondritic) and the mantle:

$$\left(\frac{^{60}Fe}{^{56}Fe}\right)_{t_{core}} = \frac{\varepsilon^{60}Ni_{mantle} - \varepsilon^{60}Ni_{chondrite}}{q_{Ni} \times f_{mantle}^{Fe/Ni}}, \quad (5)$$

where $q_{Ni} = 10^4 \left(^{56}Fe/^{60}Ni\right)_{chondrite}$ and $f_{mantle}^{Fe/Ni} = \frac{(Fe/Ni)_{mantle}}{(Fe/Ni)_{chondrite}} - 1$. To estimate the bulk $\varepsilon^{60}$Ni values of the mantle of the HED and angrite parent-bodies, we take the interpolated values from bulk rock isochrons at the inferred mantle Fe/Ni ratios (Fig. 1).

The Fe/Ni ratio of the bulk mantle of HEDs was estimated by Warren (1999) based on MgO-Ni and MgO-FeO correlations to be ~2,000 ($^{56}$Fe/$^{58}$Ni~2,700). Using this ratio and the bulk HED isochron presented in Fig. 1A, we obtain a $\varepsilon^{60}$Ni value of +0.23 ±0.13 for the bulk mantle of Vesta. Righter et al. (2008) estimated the bulk Fe/Ni ratio of the mantle of the angrite parent-body to be ~1,140 ($^{56}$Fe/$^{58}$Ni~1,500), which gives a bulk angrite mantle $\varepsilon^{60}$Ni value of +0.07 ±0.12.

The Fe/Ni ratio and $\varepsilon^{60}$Ni value of the bulk object are also needed to calculate the $^{60}$Fe/$^{56}$Fe ratio at the time of core formation. The bulk Fe/Ni ratios of Vesta and the angrite parent-body are taken to be chondritic ~17 ($^{56}$Fe/$^{58}$Ni~22; Lodders 2000; Righter 2008). A difficulty with $\varepsilon^{60}$Ni is that different chondrite groups have different $\varepsilon^{60}$Ni, $\varepsilon^{62}$Ni, and $\varepsilon^{64}$Ni values that partly correlate with isotopic anomalies in $^{54}$Cr (Dauphas et al., 2008; Regelous et al., 2008; Tang and Dauphas, 2010; Warren 2011; Steele et al., 2012). Thus, variations in $\varepsilon^{60}$Ni in bulk chondrites may be unrelated to $^{60}$Fe-decay, reflecting instead inheritance of nucleosynthetic anomalies carried by presolar phases, as had been documented previously for several other elements (Dauphas et al., 2002, 2004b; Hidaka et al., 2003; Andreasen and Sharma 2006, 2007; Trinquier et al., 2007, 2009; Carlson et al., 2007; Qin et al., 2008; Chen et al., 2010, 2011; Burkhardt et al., 2011; Moynier et al., 2012). HEDs and angrites cannot be tied to any chondrite group, so we take the average $\varepsilon^{60}$Ni value of all chondrite measurements published so far (-0.07±0.06; Dauphas et al., 2008; Regelous et al., 2008; Steele et al., 2012; Table 1).

Injecting the above parameters in Eq. 5, we calculate the $^{60}$Fe/$^{56}$Fe ratio at the time of core formation on the angrite parent-body to be $(3.46\pm3.37)\times10^{-9}$. Using a $^{60}$Fe/$^{56}$Fe initial ratio in the solar protoplanetary disk of $(11.5\pm2.6)\times10^{-9}$, this corresponds to a time of core formation of $4.5^{+11.7}_{-2.7}$ Myr after CAI [$2.7^{+11.7}_{-2.9}$ Myr if one adopts $^{60}$Fe/$^{56}$Fe=$(7.05\pm2.26)\times10^{-9}$ at CAI formation]. Because of large uncertainties in the bulk composition of the angrite parent-body, the age obtained for this object is very uncertain but is in agreement with an independent estimate from $^{182}$Hf-$^{182}$W systematics ($t_{1/2}$=8.9 Myr), indicating that the core formed within ~2 Myr of CAI formation (Kleine et al., 2012).

For Vesta, we obtain a $^{60}$Fe/$^{56}$Fe initial ratio at the time of core formation of $(4.32\pm2.08)\times10^{-9}$. This estimate is robust as the $^{60}$Ni-excess in the bulk mantle of Vesta is well resolved (Fig. 7). Using a $^{60}$Fe/$^{56}$Fe initial ratio in the solar protoplanetary disk of $(11.5\pm2.6)\times10^{-9}$, we calculate a time of core formation of $3.7^{+2.5}_{-1.7}$ Myr after CAI [$1.9^{+2.6}_{-2.0}$ Myr if one adopts $^{60}$Fe/$^{56}$Fe=$(7.05\pm2.26)\times10^{-9}$ at CAI formation]. It is the first solid estimate of the time of core formation on Vesta and is consistent with results of a thermal model of radioactive heating of Vesta by $^{26}$Al-decay suggesting a time of core formation of 4.6 Myr after CAI (Ghosh and McSween, 1998). For comparison, the $^{182}$Hf-$^{182}$W system gives a very uncertain age of 3±6 Myr for core formation in the HED parent-body (Kleine et al., 2009).

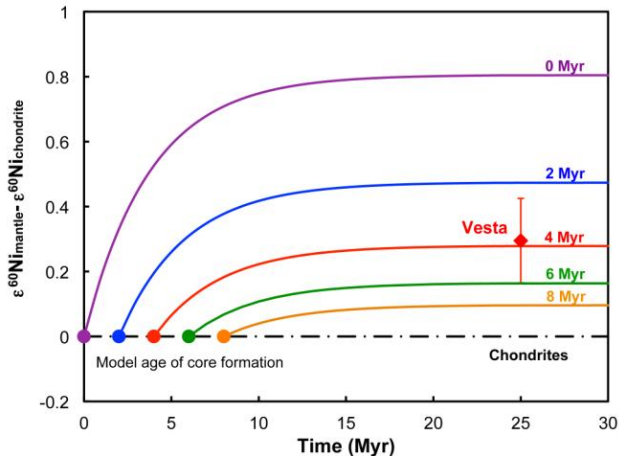

**Fig. 7.** $\varepsilon^{60}$Ni (see Fig. 1 caption for notations) isotope evolution of the mantle of Vesta ($^{58}$Fe/$^{56}$Ni=2696, Warren 1999) for different model ages of core formation. The estimated Ni isotopic composition of the bulk-mantle of HEDs ($\varepsilon^{60}$Ni$_{mantle}$-$\varepsilon^{60}$Ni$_{chondrite}$) constrains core formation on Vesta to have occurred $3.7^{+2.5}_{-1.7}$ Myr after CAIs [assuming $^{60}$Fe/$^{56}$Fe=$(1.15\pm0.26)\times10^{-8}$ at CAI formation]

## 4.3. Origins of $^{26}$Al and $^{60}$Fe in the early solar system

The new constraints presented above have far-reaching implications for the source of $^{60}$Fe in the inner solar system. The high $^{60}$Fe/$^{56}$Fe initial ratio obtained by *in situ* methods required involvement of one or several supernovae to seed the protoplanetary disk or the solar system parent molecular cloud with freshly synthesized $^{60}$Fe (Tachibana et al., 2003, 2006; Mostefaoui et al., 2004, 2005; Guan et al., 2007; Marhas and Mishra, 2012; Mishra and Chaussidon, 2012). However, the low abundance of $^{60}$Fe documented here could have simply been inherited from abundances present in the interstellar medium (ISM) that made the solar system. Analysis of γ-ray emissions from the galactic plane give a present average flux $^{60}$Fe/$^{26}$Al ratio of 0.148 ± 0.06 (Wang et al., 2007), which translates into a $^{60}$Fe/$^{26}$Al ratio of 0.148×2.62/0.705=0.55±0.22. The $^{60}$Fe/$^{56}$Fe ratio in the ISM at present can be estimated using,

$(^{60}Fe/^{56}Fe)_{ISM}=(^{60}Fe/^{26}Al)_{ISM}\times(^{26}Al/^{56}Fe)_{ISM}$.

Diehl et al. (2006) estimated the amount of $^{26}$Al in ISM gas to be 2.25±0.65 M$_\odot$. The total mass of ISM gas in the Galaxy is 4.95×10$^9$ M$_\odot$ (Robin et al., 2003). The mass fractions of $^{27}$Al and $^{56}$Fe in the protosolar nebula are X($^{27}$Al)=6.26×10$^{-5}$ and X($^{56}$Fe)=1.19×10$^{-3}$ (Lodders 2003). These fractions represent snapshots of the ISM composition 4.5 Ga. Models of the chemical evolution of the Galaxy predict that the mass fraction of a primary nuclide ($^{56}$Fe) should scale as $t$ while that of a secondary nuclide ($^{27}$Al) should scale as $t^2$ (Clayton 1985; Clayton and Pantelaki 1986; Huss et al., 2009). Correcting for this and adopting an age of ~12 Ga for the Galaxy (Dauphas, 2005 and references therein), we obtain present mass fractions in the ISM of X($^{27}$Al)=(6.26×10$^{-5}$)×(12/7.5)$^2$=1.6×10$^{-4}$ and X($^{56}$Fe)=(1.19×10$^{-3}$)×(12/7.5)=1.9×10$^{-3}$. We thus estimate present ISM atom ratios of $(^{26}Al/^{27}Al)_{ISM}$=(3.0±0.8)×10$^{-6}$ and $(^{60}Fe/^{56}Fe)_{ISM}$=(2.8±1.4)×10$^{-7}$. The ISM $^{60}$Fe/$^{56}$Fe value is ~30 times higher than the ratio measured in meteorites. According to models of galactic chemical evolution, the ratio of a secondary radioactive nuclide ($^{60}$Fe) to a primary stable nuclide ($^{56}$Fe) should be independent of the age of the disk (Huss et al., 2009). Therefore, the $^{60}$Fe/$^{56}$Fe ratio measured in the present ISM is a good proxy for the steady-state ratio at solar system formation.

Nucleosynthesis in discrete and stochastic events can lead to variations in the abundance of short-lived nuclides in the ISM. The steady-state assumption should be approximately valid for extinct radionuclides with half-lives greater than ~3-7 Myr that are produced in typical core-collapse supernovae (Meyer and Clayton, 2000). With a half-life of 2.62 Myr, $^{60}$Fe is expected to show local enhancements and depletions relative to the ISM average. In particular, the abundance of a radioactive nuclide in a star-forming molecular cloud should be lower than that of the average ISM due to partial isolation from fresh nucleosynthetic input. To address this issue, Clayton (1983) used a three phase mixing model of the ISM involving dense molecular clouds from which stellar systems form, large HI clouds, and smaller HI clouds that can be evaporated by supernova shocks. Using the same parameters as those used in Clayton (1983), the expected ratio in molecular clouds is $(^{60}Fe/^{56}Fe)_{MC}=(^{60}Fe/^{56}Fe)_{ISM}/[1+1.5T_{mix}/\tau+0.4(T_{mix}/\tau)^2]$, where $\tau=t_{1/2}/\ln(2)$ is the mean-life of the nuclide and $T_{mix}$ is the mixing timescale between the different ISM reservoirs. The calculated $^{60}$Fe/$^{56}$Fe ratio in molecular clouds can match the one measured in meteorites by assuming a mixing timescale of 15 Myr. This timescale is in agreement with that inferred from extinct $p$-process nuclides $^{92}$Nb and $^{146}$Sm (<30 Myr; Rauscher et al., submitted, supporting online material). However, it is lower than estimates based on $r$-process radionuclides (*e.g.*, $^{129}$I) that give timescales up to 300 Myr (Huss et al., 2009). This is a long-standing issue in cosmochemistry that may be related to the fact that some of these nuclides are produced in low frequency events, so the steady-state assumption is invalid (Wasserburg et al., 2006).

Although uncertain, the calculations above demonstrate that the low initial $^{60}$Fe/$^{56}$Fe ratio in the early solar system does not require direct injection from a nearby supernova. However, a nearby stellar source for $^{26}$Al is still needed. This could have been a passing AGB-star that delivered $^{26}$Al and little $^{60}$Fe through stellar winds (Wasserburg et al., 2006). However, the probability of encountering an evolved AGB-star in a star-forming region is very low (less than 3 in a million; Kastner and Myers, 1994). A massive star is therefore the favored source for $^{26}$Al, requiring special circumstances in order to avoid delivering too much $^{60}$Fe. In massive stars, $^{26}$Al is produced by proton capture on $^{25}$Mg, which occurs in more external regions than the site where $^{60}$Fe is made (Limongi and Chieffi, 2006). It is thus possible to decouple the two extinct radionuclides if

(i) $^{26}$Al was injected by stellar winds from one or several massive stars, possibly a Wolf-Rayet star (Arnould et al., 1997, 2006; Gaidos et al., 2009; Tatischeff et al., 2010; Gounelle and Meynet, 2012),

(ii) Following the explosion, only material from the external layers of the supernova were injected because the inner portion fell back onto the core (Meyer and Clayton, 2000; Takigawa et al., 2008),

(iii) Interaction of the supernova ejecta with the proto-solar cloud favored injection of the outer layers (Gritschneder et al., 2012).

We favor the first scenario of injection by winds from one or several massive stars (>30 M$_\odot$) because decoupling of $^{26}$Al from $^{60}$Fe is a natural outcome of the evolution of massive stars and such pollution is expected to occur in star-forming regions. In the parent molecular cloud to the Sun, massive stars evolved rapidly, blew-off $^{26}$Al-rich winds, and carved a bubble within the cloud. The Sun formed later and incorporated some of the bubble shell material polluted with $^{26}$Al from the first generation of massive stars.

## 5. Conclusions

We have studied the Ni and Fe isotopic compositions of meteorites to establish the abundance and distribution of $^{60}$Fe in the solar protoplanetary disk. This extinct nuclide provides critical constraints on the astrophysical context of solar system formation and on the chronology of early solar system events:

• Measurements of angrites, HEDs, and unequilibrated ordinary chondrites Semarkona and NWA 5717 reveal the presence of $^{60}$Fe ($t_{1/2}$=2.6 Myr) at a low level of (11.5±2.6)×10$^{-9}$, consistent with derivation from galactic background [$^{60}$Fe/$^{56}$Fe$_{ISM}$=(2.8±1.4)×10$^{-7}$]. At such low abundance, $^{60}$Fe was a negligible radioactive heat-source in planetary bodies.

• The main source of uncertainty in the estimation of the $^{60}$Fe/$^{56}$Fe ratio is the uncertainty that affects the initial $^{53}$Mn/$^{55}$Mn ratio at CAI formation, which is used to back-calculate the initial $^{60}$Fe/$^{56}$Fe ratio at the birth of the solar system. Taking this uncertainty into account, we obtain $^{60}$Fe/$^{56}$Fe ratios of 7×10$^{-9}$ to 12×10$^{-9}$ at CAI formation.

• Measurements of $^{58}$Fe, the most neutron-rich stable isotope of iron, demonstrate that $^{60}$Fe was uniformly distributed in the solar protoplanetary disk sampled by achondrites and chondrites. Previous work has shown that all CAIs probably did not incorporate the same amounts of $^{26}$Al. It is currently unknown whether $^{60}$Fe was uniformly distributed in CAIs, which we would predict if $^{60}$Fe was derived from galactic background.

- Aluminum-26 in the early solar system was most likely derived from stellar winds from one or several massive stars (>30 M$_\odot$). Because such winds are poor in $^{60}$Fe, this naturally explains the high abundance of $^{26}$Al relative to $^{60}$Fe in meteorites.
- We estimate the time of core formation on Vesta to be $3.7^{+2.5}_{-1.7}$ Myr after CAI formation.

**Acknowledgments:** We thank A.N. Krot, R.N. Clayton, P.R. Heck, the Robert A. Pritzker Center for Meteoritics (Field Museum) and Smithsonian National Museum of Natural History for providing the samples used in this study. Discussions with B.S. Meyer, R. Diehl, A.M. Davis, A.J. Campbell, T. Rauscher, M. Wadhwa and G.R. Huss were appreciated. Thorough reviews by A.N. Krot and two anonymous reviewers helped improve the manuscript. SELFRAG carried out high voltage pulse power fragmentation of some specimens. This work was supported by NASA (NNX09AG59G, NNX12AH60G), NSF (EAR-1144429) and a Packard Fellowship to N.D.

## Appendix A

Supplementary data to this article can be found in the online version at http://dx.doi.org/10.1016/j.epsl.2012.10.011.

## Appendix B

Let us consider 5 isotopes $i_1$, $i_2$, $i_3$, $i_4$, and $i_5$ of masses $m_1$, $m_2$, $m_3$, $m_4$, and $m_5$. The fractionating chemical species has mass $m_i + M$. To investigate the influence of mass fractionation laws on isotope measurements by internal normalization, we use the generalized power law (Maréchal et al. 1999) that relates the fractionation ratios $r$ to the true ratios $R$ through,

$$\ln(r_{2/1}) = \ln(R_{2/1}) + [(m_2 + M)^n - (m_1 + M)^n]\ln(g), \quad (B1)$$
$$\ln(r_{4/3}) = \ln(R_{4/3}) + [(m_4 + M)^n - (m_3 + M)^n]\ln(g), \quad (B2)$$

where $g$ quantifies the degree of mass fractionation and $n$ is a free parameters of mass fractionation law. To correct the ratio $r_{2/1}$ by fixing the measured ratio $r_{4/3}$ to $R_{4/3}$ by internal normalization, we have,

$$\ln(R_{2/1}) = \ln(r_{2/1}) - \frac{(m_2 + M)^n - (m_1 + M)^n}{(m_4 + M)^n - (m_3 + M)^n}\ln\left(\frac{r_{4/3}}{R_{4/3}}\right). \quad (B3)$$

If the wrong law (exponent $k$ instead of $n$) and the wrong mass ($m_i + m$ instead of $m_i + M$) had been used, the corrected ratio would be,

$$\ln(R'_{2/1}) = \ln(r_{2/1}) - \frac{(m_2 + m)^k - (m_1 + m)^k}{(m_4 + m)^k - (m_3 + m)^k}\ln\left(\frac{r_{4/3}}{R_{4/3}}\right). \quad (B4)$$

The departure on the corrected ratio from the true ratio in ε unit would therefore be (we take the difference between the previous 2 equations),

$$\varepsilon'_{2/1} \approx \left(\frac{R'_{2/1}}{R_{2/1}}-1\right)10^4 \approx \ln\left(\frac{R'_{2/1}}{R_{2/1}}\right)10^4$$

$$= \left[\frac{(m_2+M)^n - (m_1+M)^n}{(m_4+M)^n - (m_3+M)^n} - \frac{(m_2+m)^k - (m_1+m)^k}{(m_4+m)^k - (m_3+m)^k}\right]\ln\left(\frac{r_{4/3}}{R_{4/3}}\right)10^4, \quad (B5)$$

The first term in bracket can be rewritten in the form,

$$\frac{(m_2+M)^n - (m_1+M)^n}{(m_4+M)^n - (m_3+M)^n} = \frac{[(m_2+M)/(m_1+M)]^n - 1}{[(m_4+M)/(m_1+M)]^n - [(m_3+M)/(m_1+M)]^n}, \quad (B6)$$

Because the mass ratios are close to unity, the terms are well approximated by second order Taylor series expansions around 1,

$$\frac{(m_2+M)^n - (m_1+M)^n}{(m_4+M)^n - (m_3+M)^n} \approx \frac{(m_1-m_2)[m_1(n-3) + m_2(1-n) - 2M]}{(m_3-m_4)[(m_3+m_4)(1-n) + 2m_1(n-2) - 2M]}. \quad (B7)$$

Similarly we have,

$$\frac{(m_2+m)^k - (m_1+m)^k}{(m_4+m)^k - (m_3+m)^k} \approx \frac{(m_1-m_2)[m_1(k-3) + m_2(1-k) - 2m]}{(m_3-m_4)[(m_3+m_4)(1-k) + 2m_1(k-2) - 2m]}. \quad (B8)$$

By taking the difference B7-B8, it follows that Eq. B5 is well approximated by,

$$\varepsilon'_{2/1} \approx 5[n(m_1+m) - k(m_1+M) + M - m]\frac{(m_2-m_1)(m_1+m_2-m_3-m_4)}{(m_4-m_3)(m_1+m)(m_1+M)}\delta_{4/3}, \quad (B9)$$

$$\varepsilon'_{2/1} \approx 5[n(m_1+m) - k(m_1+M) + M - m]\frac{(m_2-m_1)(m_1+m_2-m_3-m_4)}{(m_1+m)(m_1+M)}F, \quad (B10)$$

where $F$ is the isotopic fractionation in ‰/amu. If another pair of isotopes 5/1 has been corrected for mass fractionation by the same procedure, we have,

$$\varepsilon'_{5/1} \approx 5[n(m_1+m) - k(m_1+M) + M - m]\frac{(m_5-m_1)(m_1+m_5-m_3-m_4)}{(m_1+m)(m_1+M)}F, \quad (B11)$$

The ratio between $\varepsilon'_{2/1}$ and $\varepsilon'_{5/1}$ is therefore,

$$\varepsilon'_{2/1} \approx \frac{(m_2-m_1)(m_1+m_2-m_3-m_4)}{(m_5-m_1)(m_1+m_5-m_3-m_4)}\varepsilon'_{5/1}. \quad (B12)$$

Often, the same isotope is used as denominator in all isotopic ratios and $m_3$ can be identified with $m_1$ (i.e., the $r_{4/1}$ ratio is used for internal normalization). In this case, Eq. B10 and B12 simplify to,

$$\varepsilon'_{2/1} \approx 5[n(m_1+m) - k(m_1+M) + M - m]\frac{(m_2-m_1)(m_2-m_4)}{(m_1+m)(m_1+M)}F, \quad (B13)$$

$$\varepsilon'_{2/1} \approx \frac{(m_2-m_1)(m_2-m_4)}{(m_5-m_1)(m_5-m_4)}\varepsilon'_{5/1}. \quad (B14)$$

If the generalized power law cannot describe the isotopic fractionation, a linear relationship between isotopic artifact (anomaly) and mass fractionation (Eq. B9 and B10) is not warranted. This is the case for the isotopic composition of the product of a

Rayleigh distillation (*e.g.*, solid during gas condensation), in which case the relationship between ε and δ is best investigated using a numerical approach. However, Eq. B10 is still applicable for the residue of a distillation. Let us consider a distillation with a fractionation factor of $\alpha_{2/1} = (m_1/m_2)^\beta$. The isotopic composition in the residue following such a distillation is,

$$\ln(r_{2/1}) = \ln(R_{2/1}) + [(m_1/m_2)^\beta - 1]\ln(f), \quad \text{(B15)}$$
$$\ln(r_{4/3}) = \ln(R_{4/3}) + [(m_3/m_4)^\beta - 1]\ln(f), \quad \text{(B16)}$$

where $f$ is the fraction of $i_1$ left in the residue. The internally corrected $i_2/i_1$ ratio is,

$$\ln(R_{2/1}) = \ln(r_{2/1}) - \left[\frac{(m_1/m_2)^\beta - 1}{(m_3/m_4)^\beta - 1}\right] \ln\left(\frac{r_{4/3}}{R_{4/3}}\right). \quad \text{(B17)}$$

As long as $(m_3/m_1)^\beta \approx 1$,

$$\ln(R_{2/1}) \approx \ln(r_{2/1}) - \left[\frac{(m_1/m_2)^\beta - 1}{(m_1/m_4)^\beta - (m_1/m_3)^\beta}\right] \ln\left(\frac{r_{4/3}}{R_{4/3}}\right), \quad \text{(B18)}$$

$$\ln(R_{2/1}) \approx \ln(r_{2/1}) - \left[\frac{m_2^{-\beta} - m_1^{-\beta}}{m_4^{-\beta} - m_3^{-\beta}}\right] \ln\left(\frac{r_{4/3}}{R_{4/3}}\right), \quad \text{(B19)}$$

By identification with Eq. (B3), it follows $n = -\beta$ and Eq. B10 applies. Again, this is only valid for the residue of a distillation with $\alpha_{2/1} = (m_1/m_2)^\beta$. Note that $\varepsilon'_{2/1}$ does not follow mass-balance between product and residue; $\varepsilon'_{2/1,\ residue}\, f + \varepsilon'_{2/1,\ product}(1-f) \neq 0$.